\documentclass[a4paper,11pt]{article}
\usepackage[pdftex]{graphicx}
\usepackage[T1]{fontenc}
\usepackage{lmodern}
\usepackage{slashed}
\usepackage[utf8]{inputenc}
\usepackage[english]{babel}
\usepackage{microtype}
\usepackage{cite}
\usepackage{amsmath}
\usepackage{amssymb}
\usepackage{amsfonts}
\usepackage[nottoc,notlot,notlof]{tocbibind}
\usepackage{upgreek}
\usepackage{mathtools}
\numberwithin{equation}{section}
\allowdisplaybreaks
\usepackage[all]{xy}
\usepackage{color}
\usepackage{graphicx}
\graphicspath{{images/}}
\usepackage{geometry}
\usepackage[toc,page]{appendix}
\usepackage{hyperref}
\usepackage[normalem]{ulem}

\usepackage{bm}
\usepackage{ragged2e}
\usepackage{appendix}
\usepackage{slashed}
\usepackage{bbold}
\usepackage{cancel}

\definecolor{blue-violet}{rgb}{0.54, 0.17, 0.89}
\definecolor{PineGreen}{cmyk}{0.92, 0, 0.59, 0.25}
\definecolor{YellowOrange}{cmyk}{0, 0.42, 1, 0}
\definecolor{orange}{rgb}{0.95, 0.5, 0.1}

\DeclareMathAlphabet{\mathpzc}{OT1}{pzc}{m}{it}

\usepackage[dvipsnames]{xcolor}

\def\noi{\noindent}

\def\al{\alpha}

\def\ga{\gamma}

\let \part\partial

\def\Dcal{{\cal D}}

\def\epsilonbar{{\bar \epsilon}}
\def\thetabar{{\bar \theta}}
\def\psibar{\bar \psi}
\def\epsilonbar{\bar \epsilon}

\def\rhobar{\bar \rho}

\def\epsi{\varepsilon}

\newcommand{\eq}{\begin{equation}}
\newcommand{\eqa}{\begin{eqnarray}}  
\newcommand{\en}{\end{equation}}
\newcommand{\ena}{\end{eqnarray}}
\newcommand{\enn}{\nonumber \end{equation}}

\begin{document}

\begin{titlepage}

\vskip 2em
\begin{center}
{\Huge \bf Hodge Duality and Supergravity} \\[3em]

\vskip 0.5cm

{\bf
Leonardo Castellani and Pietro Antonio Grassi}
\medskip

\vskip 0.5cm

{\sl Dipartimento di Scienze e Innovazione Tecnologica
\\Universit\`a del Piemonte Orientale, viale T. Michel 11, 15121 Alessandria, Italy\\ [.5em] INFN, Sezione di 
Torino, via P. Giuria 1, 10125 Torino, Italy\\ [.5em]
Regge Center for Algebra, Geometry and Theoretical Physics, via P. Giuria 1, 10125 Torino, Italy
}\\ [4em]
\end{center}

\begin{abstract}

The Hodge dual operator, recently introduced for supermanifolds, is used to reformulate 
 super Yang-Mills and supergravity in $D=4$. We first recall the definition of the Hodge dual operator 
 for flat and curved supermanifolds. Then we show how to recover the usual super-Yang-Mills 
 equations of motion for $N=1,2$ supersymmetry, and the obstacles (as seen from Hodge dual point of view) in the case $N \geq 3$. We reconsider several ingredients of supergeometry, relevant for a superspace formulation of supergravity, in terms of the Hodge dual operator. Finally we discuss how $D=4$ and $N=1$ supergravity is obtained in this framework. 
 
\end{abstract}

\vskip 9cm\
 \noi \hrule \vskip .2cm \noi {\small
leonardo.castellani@uniupo.it, pietro.grassi@uniupo.it}

\end{titlepage}

\newpage
\setcounter{page}{1}

\tableofcontents

\section{Introduction}

Since its early days, supergravity \cite{FFvN,DZ,PvNreport,FVP} has been formulated using superspace \cite{GGRS,WB} and supergroup manifold \cite{gm11,gm12,book,DFTvN,gm24,RD1,LC1} techniques. In both these approaches local supersymmetry arises as invariance of the action under superdiffeomorphisms. While the superspace actions involve {\sl superfields}, whose $\theta$ (anticommuting coordinates) expansions contain the physical and auxiliary fields, the group manifold actions are given directly in terms of these fields, each of them originating from the {\sl supergroup manifold vielbein}. 

In the superspace approach, supergravity actions are constructed as integrals on superspace, and actions on $d$-dimensional spacetime are obtained after Berezin integration on the $\theta$ coordinates. 

In the supergroup manifold approach, the action is defined as an integral on a $d$-dimensional bosonic {\sl submanifold} of the supergroup manifold. The Lagrangian is a $d$-superform, written in terms of the supergroup vielbein and its exterior derivative. 

A theory of integration on supermanifolds in the language of differential forms has been recently developed in ref.s \cite{Castellani:2014goa,Castellani:2015paa,Castellani:2015ata,Castellani:2016ibp,CCG} , and makes use of {\sl integral forms}, the analogue of top forms for supermanifolds. Integral forms are instrumental for a rigorous definition of an action principle, and have been applied in the last years to supersymmetric and supergravity models \cite{Castellani:2014goa,Castellani:2015paa,Castellani:2015ata,Castellani:2016ibp,CCG}. Only integral forms can be integrated on supermanifolds, and therefore the $d$-superform Lagrangian must be converted into an integral form. This can be done by multiplying it with the analogue of the Poincaré dual 
(as in ordinary integration on submanifolds), called Picture Changing Operator (PCO), a representative of a De Rahm cohomology class. It is characterized by a zero form number and a ``picture number" equal to the fermionic dimension of the supermanifold, as explained in Appendix A. If there exists only one such cohomology class, one can still choose different representatives. For each choice, the final form of the action will have a different expression, but all these expressions are equivalent and correspond to the same field theory. In the case of more than one cohomology class, one can construct actions describing physically different theories, each corresponding to a particular cohomology class.

Another way to construct integral forms is to combine pieces with different form and picture numbers, without reference to a $d$-superform Lagrangian. This second way often makes use of a generalization of the Hodge dual for supermanifolds, introduced in ref. \cite{Castellani:2015ata,CCGir} by means of a generalized Fourier transform.

In this paper we obtain two main results: i) the extension to $d=4$ of the integral form construction 
for supergravity, and the proof of its equivalence with the superspace formulation, with the same logic used in ref. \cite{Castellani:2016ibp} for the $d=3$ case, i.e. with the use of appropriate PCO's; ii)  the reformulation of super Yang-Mills and supergravity theories in $d=4$ by using the super Hodge dual of ref.s \cite{Castellani:2015ata,CCGir}.

The paper is organized as follows. In Section 2 we use Cartan calculus and the super Hodge dual to rewrite cumbersome formulas in superspace geometry \cite{GGRS,Kuzenko:1998tsq} in a simple and transparent way. We show how the Berezinian emerges 
naturally and what is his relation with chiral volume forms. Super Yang-Mills and supergravity in $d=4$ are 
reformulated using the super Hodge dual. In Section 3 we give the group manifold treatment of $d=4$ supergravity with the ``old minimal'' set \cite{FPvN} of auxiliary fields. The resulting 4-form Lagrangian is then multiplied by a suitably chosen PCO, thus obtaining an integral form that can be integrated on superspace. The result is shown to coincide with the superspace expression of the action as given in \cite{GGRS}, i.e. the superspace integral of the superdeterminant. A similar exercise is carried out in Section 4,
where the ``new minimal'' set of auxiliary fields \cite{SW} is employed. Finally, in the Appendices we
discuss in more detail aspects of supermanifold geometry, and give the derivation of some useful formulas used in the text.

\section{Field Theories and Hodge Dual Operator}

In this Section, we translate some of the superspace equations in terms of the super Hodge dual introduced 
in \cite{CCGir,Castellani:2015ata}. 
We use here the following notation: indices $A,B, \dots$ refer to the
super tangent space: $A = (a, \alpha), B= (b, \beta)$, where $a,b, \dots =0,\dots,3$ are the tangent vector 
indices and $\alpha, \beta, \dots=1,\dots,4$ the spinorial tangent indices. We denote by $M,N, \dots$ 
the curved indices of the supermanifold $M = (m, \mu), N= (n, \nu)$ where $m,n, \dots$ are the 
indices of the bosonic coordinates and $\mu,\nu, \dots$ the indices of the fermionic coordinates. 
Collectively, we denote by $Z^M =(x^m, \theta^\mu)$ the coordinates of the supermanifold. 
We denote by $\mathbb{R}^{(4|4)}$ the superspace (flat supermanifold). In appendix A  we collect the definitions for forms on supermanifolds and integration theory; 
for more details, we refer to the literature \cite{Witten:2012bg,Castellani:2015paa,Noja}. 

After fixing the geometrical setting, and discussing some properties of the super Hodge dual operator $\star$,  we present in this Section simple applications to scalar field theory, abelian gauge theory and supergravity. 
We will use uniquely the Hodge dual operator, without referring to the group manifold (or {\it rheonomic}) approach \cite{book} (see \cite{LC1,RD1} for recent reviews), where the action is constructed \cite{Castellani:2014goa,Castellani:2015paa,Castellani:2016ibp}  from the rheonomic Lagrangian as follows 
\begin{eqnarray}
\label{SGA}
S = \int_{\mathcal{SM}} \mathcal{L}_{rheo}^{(4|0)} \wedge \mathbb{Y}^{(0|4)}
\end{eqnarray}
The suffix $(4|0)$ stands for form number equal to four and 
picture number equal to zero (as defined in Appendix A). $ \mathcal{L}_{rheo}^{(4|0)}$ is the rheonomic Lagrangian written in terms 
of the form fields and their differentials, while $ \mathbb{Y}^{(0|4)}$ is the Picture Changing Operator (PCO) converting the action 
from a $(4|0)$ form to a $(4|4)$ form, which can be integrated on the supermanifold $\mathcal{SM}$. 
The action $S$ depends on the supergravity fields in the rheonomic Lagrangian $ \mathcal{L}_{rheo}^{(4|0)}$ and in the  PCO $ \mathbb{Y}^{(0|4)}$, and being integrated on the entire supermanifold $\mathcal{SM}$ is automatically invariant under super-reparametrizations. In this framework the action is built {\sl without} using the Hodge dual operator $\star$. 

In the present Section we adopt a different strategy: we replace the factorized form of the action with a novel one
\begin{eqnarray}
\label{SGB}
S = \int_{\mathcal{SM}} \mathcal{L}_{\star}^{(4|4)}
\end{eqnarray}
where a suitable Hodge dual operator $\star$ is used. The Lagrangian is no longer factorized into the wedge product of a $(4|0)$-form with a $(0|4)$ form.  
Again, since $\mathcal{L}_{\star}^{(4|4)}$ is a $(4|4)$ form integrated on the 
entire supermanifold $\mathcal{SM}$, it is invariant under super-reparametrizations. 

Consider the actions for (free) scalar field theory and 
abelian gauge theory:  
\begin{eqnarray}
\label{fA}
S[\phi] = \int_{\mathcal{M}} d\phi \wedge \star d\phi\,, ~~~~~~
S[A] = \int_{\mathcal{M}} d A\wedge \star d A\,,
\end{eqnarray}
where $\mathcal{M}$ is a bosonic manifold, $\phi$ is a scalar field and $A$ is $1$-form potential. 
The operator $\star$ is the Hodge dual operator related to a metric $g$ on the manifold $\mathcal{M}$ and 
the actions \eqref{fA}, which are $4$-forms integrated on a 4-dimensional manifold, are manifestly reparametrization invariant.  In the following, 
we replace the manifold $\mathcal{M}$ with a supermanifold $\mathcal{SM}$, the $\star$ operator with a corresponding 
super $\star$ operator (see \cite{Castellani:2015ata,CCGir} for a general treatment); $\phi$ is replaced by a superfield at zero picture and $A$ is replaced by a $1$-superform at zero picture (the picture number being defined in Appendix A).

\subsection{Super Hodge dual operator}

We define the super Hodge dual operator $\star$ as follows. 
Given a form $\omega(x,\theta,V, \psi)$, considered as 
a generalized function of the supervielbein 
$(V^a, \psi^\alpha)$ and of the coordinates $(x^a, \theta^{\alpha})$ (see Appendix A),  
its Hodge dual is written in terms of a Fourier transform 
\begin{eqnarray}
\label{HDOA}
\star \omega(x,\theta,V, \psi) = i^{r^{2} -n^{2}} i^{l} \int_{\mathcal{SM}'} e^{i (\nu^a \eta_{ab} V^b + p^\alpha 
\lambda C_{\alpha\beta} \psi^\beta)} \omega(x,\theta,\nu, p) [d^4\nu d^4p]
\end{eqnarray}
where $r$ is the number of $V$'s, $l$ is the number of $\psi$'s, $n$ is the bosonic dimension,
and $\mathcal{SM}'$ is the dual superspace whose fundamental coordinates  
are $(\nu^a, p^\alpha)$ (respectively, anticommuting and commuting). The 
symbol $[d^4\nu d^4p]$ denotes the Berezin integral over $\nu^a$ and the 
Riemann-Lebesgue integral over $p^\alpha$. The metric $g$ to which $\star$ is related is 
given by the tensor\footnote{Note that, since $V^a$ and $\psi^\alpha$ are respectively bosonic and fermionic 1-forms, $\psi^\alpha \wedge \psi^\beta = \psi^\beta \wedge \psi^\alpha$ and 
$V^a \wedge V^b = - V^b \wedge V^a$.}

 \begin{eqnarray}
\label{HDOAB}
g = \eta_{ab} V^a \otimes V^b + \lambda C_{\alpha\beta} \psi^\alpha \otimes \psi^\beta
\end{eqnarray}
where $\lambda$ is a dimensionful constant.  This constant is needed in order to 
preserve homogeneity of dimensions between the two terms since $V^a$ scales as the square of  
$\psi^{\alpha}$. For $\lambda\rightarrow 0$ the metric is degenerate. $C_{\alpha\beta}$ is the charge conjugation antisymmetric matrix (this holds specifically for the
case $D=4$). The function $\omega(x, \theta, \nu, p)$ is the function obtained from $\omega(x,\theta,V, \psi)$ by substituting  $V \rightarrow \nu$ and $\psi \rightarrow p$, leaving the supermanifold coordinates $x,\theta$ untouched.

We can easily compute the Hodge dual of the supervielbeins 
\begin{eqnarray}
\label{HDOB}
\star V^a =  V^{a}_3 \delta^4(\psi)\,, ~~~~
\star \psi^\alpha = \lambda^{-1} C^{\alpha\beta} V^4 \iota_\beta \delta^4(\psi)\,. 
\end{eqnarray}
where 
\begin{eqnarray}
\label{HDOBA}
V^{ab}_{2} &\equiv& 
\epsilon^{abcd} \frac{1}{2!} \eta_{cc'} \eta_{dd'} \wedge V^{c'} \wedge V^{d'}\,, ~~~~~~
V^{a}_3 \equiv\epsilon^{abcd} \frac{1}{3!}\eta_{bb'} \eta_{cc'} \eta_{dd'} V^{b'}\wedge V^{c'} \wedge V^{d'}\nonumber \\
V^4 = V_4 &\equiv&\frac{1}{4!} \epsilon_{abcd} V^a\wedge V^b\wedge V^c \wedge V^d\,, ~~~~~~
\delta^{4}(\psi) \equiv \frac{1}{4!}\epsilon_{\alpha\beta\gamma\delta} \delta(\psi^{\alpha}) \delta(\psi^{\beta}) \delta(\psi^{\gamma}) \delta(\psi^{\delta}) \nonumber\\  
\iota_{\alpha} \delta^{4}(\psi) &\equiv& \frac{\partial}{\partial \psi^{\alpha}}  \delta^{4}(\psi)\,,~~~~~~~~~~~~~~~~~~~~~~~~
{\rm Vol}^{(4|4)} \equiv V^4 \delta^4(\psi)
\,. 
\end{eqnarray}
and the pseudoform $\delta(\psi^{\alpha})$ (see Appendix A) naturally arises as the Fourier transform $\delta(\psi^{\alpha})  = \int d{p} ~e^{i p \psi^{\alpha}}$. 
Notice that the scaling dimensions of $V^{a}$ and of its dual $\star V^{a}$ are the same, as well as for $\psi^{\alpha}$ and $\star \psi^{\alpha}$. 
Using the above definitions, we immediately find 
\begin{eqnarray}
\label{HDOC}
V^a \wedge \star V^b = \eta^{ab} {\rm Vol}^{(4|4)}\,, ~~~
\psi^\alpha \wedge \star \psi^\beta = \lambda^{-1} C^{\alpha \beta} {\rm Vol}^{(4|4)}\,, ~~~
\star {\rm Vol}^{(4|4)}  = 1\,, ~~~
\star 1 =  {\rm Vol}^{(4|4)}\,.
\end{eqnarray}
With our conventions, the dimension of ${\rm Vol}^{(4|4)}$ is equal to $2$. 

As well known, the Hodge dual operator for usual manifolds is a key ingredient for defining the 
Laplace-Beltrami operator acting on differential forms. In the present context we 
can consistently extend the definition of the Laplace-Beltrami operator to supermanifolds as 
\begin{eqnarray}
\label{LBa}
\Delta = d^{\dagger} d + d d^{\dagger} \equiv \star d \star d + d \star d \star \equiv  d^{\dagger} d + d d^{\dagger}
\end{eqnarray}
the conjugated differential operator being defined by $d^{\dagger} \equiv \star ~d~ \star$. Note that we can define a 
pairing between two forms on a supermanifold, with complementary form numbers $p, 4-p$ and complementary picture numbers $q, 4-q$,  as 
\begin{eqnarray}
\label{LBb}
\langle \omega, \eta \rangle = \int_{\mathcal{SM}} \omega\wedge \star \eta 
\end{eqnarray}
and consequently $\langle \omega, d \eta \rangle = \langle d^{\dagger}\omega, \eta \rangle$. 
This pairing does {\it not} define a positive definite scalar product on forms due to the symplectic nature of the 
scalar product for the fermionic components of a vector field. Because of that, 
even though we have the identity $\langle \omega, \Delta \omega\rangle = \langle d\omega, d\omega\rangle + 
\langle d^{\dagger}\omega, d^{\dagger}\omega \rangle$ based on Stokes' theorem for 
supermanifolds, we cannot conclude that $d\omega =0$ and 
$d^{\dagger} \omega =0$ if $\Delta \omega =0$. The Hodge theory for supermanifolds will be discussed in 
\cite{RG}.

We can compute the Laplacian \eqref{LBa} on the simplest examples of a superfield $\phi$ and of a top-integral form $\phi^{(4|4)}$. 
On a zero form $\phi$ we find 
\begin{eqnarray}
\label{HDOCA}
d \phi &=& V^a \nabla_a \phi + \psi^\alpha \nabla_\alpha \phi\nonumber \\
\star d \phi &=& V^{a}_3 \delta^4(\psi) \nabla_a \phi + 
  \lambda^{{-1}} V^4 C^{\alpha\beta} \iota_\alpha \delta^4(\psi) \nabla_\beta \phi\,. \nonumber \\
  d (\star \phi) &=& d (\phi {\rm Vol}^{(4|4)}) = 0\,, \nonumber \\
  \Delta \phi &=& (\star d \star d + d \star d \star)\phi = 
  (\eta^{ab} \nabla_a \nabla_b + \lambda^{{-1}} C^{\alpha\beta} \nabla_{\alpha} \nabla_{\beta} ) \phi 
 \end{eqnarray}
 where $\nabla_{a}$ = $E^{M}_{a} \partial_{M}$, 
 $\nabla_{\alpha} = E^{M}_{\alpha} \partial_{M}$, $E^{M}_{A}$ is the supervielbein and 
 $\partial_{M} = (\partial_{m}, \partial_{\mu})$. 
$\Delta$ is the generalized Laplacian on superfields and the two terms of the last equation in \eqref{HDOCA} 
scale with different powers of $\lambda$. 

In the same way, considering a top integral form $\phi^{(4|4)} = \phi {\rm Vol}^{(4|4)}$, where $\phi$ is its 
section, we have 
\begin{eqnarray}
\label{HDOCC}
\Delta \phi^{(4|4)} =  (\star d\! \star d  + d \star \!d \star) ( \phi {\rm Vol}^{(4|4)}) =
\left[ (\eta^{ab} \nabla_a \nabla_b + \lambda^{{-1}} C^{\alpha\beta} \nabla_\alpha \nabla_{\beta}) 
 \phi \right] {\rm Vol}^{(4|4)}
\end{eqnarray}
which is again the Laplacian operator on the section $\phi$ of the 
integral form $\phi^{(4|4)}$, multiplied by ${\rm Vol}^{(4|4)}$.  

\subsection{Chiral Superfields}

The smallest irreducible representation of supersymmetry with scalar bosonic degrees of freedom is described by a {\it chiral superfield} (see the textbooks \cite{WB,GGRS,Kuzenko:1998tsq}). 
The components of an off-shell chiral superfield describe a complex scalar, its fermionic superpartner and a complex auxiliary field, i.e. 4 (2)  bosonic and 4 (2) fermionic degrees of freedom if off-shell (on shell). A superfield 
$\phi$ is in general a reducibile representation: imposing some invariant equations we can single out 
irreducible representations. This is done by setting $D_{L} \phi=0$ and $D_{R} \bar\phi=0$ for chiral $\phi$ and 
anti-chiral ${\bar \phi}$ superfields, where $(D_L)_{\alpha} = \frac12 (1+\gamma^{5})_{\alpha}^{~\beta} \nabla_{\beta}$ and 
$(D_{R})_{\alpha} = \frac12 (1-\gamma^{5})_{\alpha}^{~\beta} \nabla_{\beta}$. 
In this subsection, we illustrate how a chiral superfield can be defined
using the Hodge dual operator $\star$ and its chiral/anti-chiral relatives $\star_{C}, \star_{\bar C}$. 

In the case of flat 4D superspace, one can define a {\it chiral} and an {\it anti-chiral volume} form as follows 
\begin{eqnarray}
\label{CVOFA}
{\rm Vol}^{(4|2)}_{L}= \epsilon_{abcd} V^a\dots V^d  \delta^{2}(\psi_{L})\,, ~~~~
{\rm Vol}^{(4|2)}_{R} = \epsilon_{abcd} V^a\dots V^d  \delta^{2}(\psi_{R})\,, ~~~~
\end{eqnarray}
where $\psi = \psi_{L} + \psi_{R}$ and $\psi_{L/R} = (1\pm \gamma^{5})/2 \psi$. 
The volume forms ${\rm Vol}_{L}$ 
and ${\rm Vol}_{R}$ have form degree equal to 4, but picture number equal to $2$ unlike ${\rm Vol}^{{(4|4)}}$. 
They transform under the chiral transformations as 
follows 
\begin{eqnarray}
\label{CVOFB}
{\rm Vol}^{'(4|2)}_{L} = {\rm Sdet}_L(J)^{-1} {\rm Vol}^{(4|2)}_{L} \,, ~~~~
{\rm Vol}^{'(4|2)}_{R} = {\rm Sdet}_{R}(\bar J)^{-1} {\rm Vol}^{(4|2)}_{R} \,, ~~~~
\end{eqnarray}
where $J$ ($\bar J$)  is a $6\times 6$ supermatrix corresponding to coordinate transformations in chiral (antichiral) superspace. This means that the 
coordinate transformations are generated by chiral/anti-chiral superfields as we now describe. 
Notice that although ${\rm Vol}^{(4|2)}_{L}$ 
and ${\rm Vol}^{(4|2)}_{R}$ are not top integral forms, they are closed, i.e. $ d {\rm Vol}^{(4|2)}_{L} =0$ 
and $d {\rm Vol}^{(4|2)}_{R} =0$, since $d V^{a} = \bar\psi_{L} \gamma^{a} \psi_{R}$ and either $\psi_{L}$ or 
$\pi_{R}$ are annihilated by $\delta^{2}(\psi_{L})$ or $\delta^{2}(\psi_{R})$. 
The volume form ${\rm Vol}^{(4|4)}$ is related to the chiral volume forms as 
\begin{eqnarray}
\label{volPCO} 
{\rm Vol}^{(4|4)} = {\mathcal R}_{L} {\rm Vol}^{(4|2)}_{L}\wedge \mathbb{Y}_{R} +  
 {\mathcal R}_{R} {\rm Vol}^{(4|2)}_{R}\wedge \mathbb{Y}_{L}
\end{eqnarray}
where $ \mathbb{Y}_{L}$ and $ \mathbb{Y}_{R}$ are suitable PCO's and 
where   ${\mathcal R}_{L},{\mathcal R}_{R}$ are two superfields. 
We will discuss further this point in the context of a generic supermanifold; it leads to Siegel's formula for the volume of a $D=4, N=1$ supermanifold. 

Now we can define {\it chiral} and {\it anti-chiral superfields} as follows 
\begin{eqnarray}
\label{CVOFC}
d\phi \wedge {\rm Vol}^{(4|2)}_{L}= 0\,, ~~~~~ 
d\bar \phi \wedge {\rm Vol}^{(4|2)}_{R} = 0\,, ~~~~~
\end{eqnarray}
Notice that $d\phi=V^a \partial_a \phi+{\bar \psi_R} D_R \phi + {\bar \psi_L} D_L \phi$ is a $1$-superform, and therefore the above equations select only the following pieces 
\begin{eqnarray}
\label{CVD}
\bar\psi_{R} D_{R}\phi \wedge {\rm Vol}^{(4|2)}_{L}=0\,, ~~~~~~
\bar\psi_{L} D_{L}\bar\phi \wedge {\rm Vol}^{(4|2)}_{R}=0\,, 
 \end{eqnarray}
since the remaining terms are set to zero either by the product of $V$'s or by $\delta^{2}(\psi_{L/R})$. 
Therefore, by imposing \eqref{CVD}, we obtain the usual chiral and anti-chiral constraints 
$D_{R}\phi =0$ and $D_{L}\bar\phi=0$.\footnote{In the group manifold framework this is obtained by choosing the 
differential of $\phi$ as $d \phi = V^{a} \nabla_{a} \phi + \psi_{L} W_{L}$ where $W_{R}$ is set to zero from the 
beginning \cite{book}.}

We introduce the chiral and anti-chiral Hodge dual operators (defined in the same way as 
the full Hodge dual operator \eqref{HDOA}, where the fermionic variables are replaced by their chiral or antichiral counterparts), denoted by $\star_C$ and $\star_{\bar C}$. Then we have 
\begin{eqnarray}
\label{CVE}
\star_L \phi = \phi\, {\rm Vol}^{(4|2)}_L\,, ~~~~~
\star_{R} \bar\phi = \bar\phi \, {\rm Vol}^{(4|2)}_{R}\,, ~~~~~
\star_{L} 1 =  {\rm Vol}^{(4|2)}_L\,, ~~~~~
\star_{R} 1 =  {\rm Vol}^{(4|2)}_{R}\,,
\end{eqnarray}
Acting with $d$ yields 
\begin{eqnarray}
\label{CVEA}
d (\star_L \phi) = d\phi \wedge\,  {\rm Vol}^{(4|2)}_{L} \,,  ~~~~~
d (\star_{R} \bar\phi) = d\bar\phi \wedge \, {\rm Vol}^{(4|2)}_{R} \,,  ~~~~~
\end{eqnarray}
Acting again with $\star_{L}$ and $\star_{R}$ on both members of these equations, we find
\begin{eqnarray}
\label{CVF}
\star_{L} d (\star_L \phi) = \bar\psi_{R} D_{R} \bar\phi \,, ~~~~~~
\star_{R} d (\star_{R}\bar\phi) = \bar\psi_{L} D_{L}\bar\phi\,.
\end{eqnarray}
and therefore the chirality conditions are equivalent to
\begin{eqnarray}
\label{CVFA}
d^{\dagger}_{L} \phi =0\,, ~~~~~~
d^{\dagger}_{R} \bar \phi =0\,. 
\end{eqnarray}
For a $0$-form, we have $d^{\dagger} \phi =0$ because $d^{\dagger}: \Omega^{(p|q)} \rightarrow 
\Omega^{{(p-1|q)}}$. This is due to the fact that $\star \phi$ is a top integral form and 
therefore $d (\star \phi)$ automatically vanishes. When using chiral Hodge dual operators $\star_{L}$ 
or $\star_{R}$ this is no longer true in general, but remains true on chiral/anti-chiral superfields. With this language the chirality conditions have a clear geometrical origin.

If we compute the Laplace-Beltrami differential for a chiral/anti-chiral superfield 
we get 
\begin{eqnarray}
\label{CFVB}
\Delta_{C} \phi &=& ( d^{\dagger}_{C} d + d d^{\dagger}_{C }) \phi = d^{\dagger}_{C} d\phi = 
(\eta^{ab} \nabla_{a}\nabla_{b} + \lambda^{-1} D_{L}^{2}) \phi\,, ~~~~ \nonumber \\
\Delta_{\bar C} \bar\phi &=& ( d^{\dagger}_{\bar C} d + d d^{\dagger}_{\bar C }) \bar\phi = 
d^{\dagger}_{\bar C} d\bar\phi = 
(\eta^{ab} \nabla_{a}\nabla_{b} + \lambda^{-1} D_{R}^{2}) \phi\,,
\end{eqnarray}
where $D_{L}^{2} = \bar D_{L} D_{L}$ and $D_{R}^{2} = \bar D_{R} D_{R}$.  
The free equations of motion $\Delta_{C} \phi=0$ and $\Delta_{\bar C} \bar\phi =0$ 
for chiral/anti-chiral superfields are obtained in the limit $\lambda \rightarrow 0$. 
Otherwise, these equations are higher-derivative modifications of the free equations of motion. 

Before discussing the action, we consider another type of multiplet. As is known, there is a second irreducible multiplet with scalar bosonic degrees of freedom: the {\it linear} superfield. It is 
defined to satisfy $D^{2}_{R} \phi=0$ and $D^{2}_{L} \bar\phi=0$ and the difference with the chiral/anti-chiral 
superfields is the different set of auxiliary fields for the off-shell multiplet. 
From the computation \eqref{HDOCA}, we see that 
\begin{eqnarray}
\label{CFVC}
(d^{\dagger} - d^{\dagger}_{C})d \phi &=& \lambda^{-1} D^{2}_{R} \phi \nonumber \\
(d^{\dagger} - d^{\dagger}_{\bar C})d \bar\phi &=& \lambda^{-1} D^{2}_{L} \bar\phi
\end{eqnarray}
and therefore the requirement $D^{2}_{R} \phi=0$ and $D^{2}_{L} \bar\phi=0$ 
implies that on $d\phi$ the 
two differentials $d^{\dagger}$ and $d^{\dagger}_{C}$ coincide. Since the linear superfields and 
the chiral superfields are related by duality transformations (see \cite{GGRS}), it would be interesting to
verify whether
indeed the geometric equation \eqref{CFVC} emerges as dual to \eqref{CVFA}.  
 
The action can be written using the Hodge dual operator $\star$ 
for a chiral super field ${\rm Vol}_{R} \wedge d\phi =0$ as follows 
\begin{eqnarray}
\label{sscaC}
S = \int_{\mathcal{SM}^{{(4|4)}}} \star \bar \phi \phi
\end{eqnarray}
which reduces to the usual chiral superfield action by integrating on the 
cotangent directions $V^{a},\psi_{L}$ and $\psi_{R}$. Notice that 
a geometrical action $S = \int_{\mathcal{SM}^{{(4|4)}}}  d\bar \phi \wedge \star d\phi$ 
would yield a high derivative theory after Berezin integration.

\subsection{Gauge Fields}

On a bosonic manifold $\mathcal{M}$, the Bianchi identity and the Maxwell equations for the 
abelian potential $A$ are given by 
\begin{eqnarray}
\label{GTA}
d F =0\,,  ~~~~~~ d^{\dagger} F = J 
\end{eqnarray}
where $J$ is the conserved ($d^{\dagger} J=0$) 
electric current coupled to the gauge field $A$ and $F = d A$ is its field strength.  

Moving to a supermanifold,  one considers a gauge superfield $A$ and 
its field strength $F$ which are superforms 
\begin{eqnarray}
\label{GTAA}
A = A_{a} V^{a} + A_{\alpha} \psi^{\alpha}\,, ~~~~
F = F_{ab} V^{a} V^{b} + F_{a\beta} V^{a} \psi^{\beta} + F_{\alpha \beta} \psi^{\alpha} \psi^{\beta} \,.
\end{eqnarray}
They have many components since $A_{a}, A_{\alpha}$ are superfields. In order to single out the physical ones, one imposes the condition that $F_{\alpha\beta} =0$ and by means of the 
Bianchi identity $dF=0$, we find that there are only one gauge vector field, the gaugino and 
one auxiliary field. Thus, the field strength can be written as 
\begin{eqnarray}
\label{GTAB}
F = f_{ab} V^{a} V^{b} +   \bar W \gamma_{a} \psi V^{a}
\end{eqnarray}
where $W^{\alpha}$ is the gaugino field strength, and $\gamma^{a}$ are the Dirac matrices in the Majorana 
representation.  In the component expansion of $W^{\alpha}$ (as illustrated 
in \cite{GGRS} or \cite{WB}) one retrieves the physical degrees of freedom and the auxiliary fields. 
The gauge field strength $f_{ab}$ is not an independent superfield since it satisfies 
$f_{\alpha\beta} = \frac14 \nabla \gamma_{ab} W$ as a consequence of the Bianchi identity, 
where $\nabla_{\alpha}$ is the superderivative introduced in the previous section.  
Again, as a consequence of the Bianchi identities, we have $\nabla_{\alpha} W^{\alpha} =0$.

Computing the Hodge dual of $F$ we obtain
\begin{eqnarray}
\label{gauB}
\star F = f_{ab} \eta^{aa'} \eta^{bb'} 
\epsilon_{a'b'cd} V^c V^d \delta^4(\psi) +  \lambda^{-1}
(V_3)^a \bar W \gamma_a C\iota \delta^4(\psi) 
\end{eqnarray}
Notice that the charge conjugation matrix $C$ enters because, due to the Hodge dual operator, indices are contracted covariantly. Then we compute the differential and 
finally the Hodge dual again, finding
\begin{eqnarray}
\label{gauBA}
\star d \!\star F \equiv d^{\dagger} F =  (\nabla^a f_{ab} - \lambda^{-1} \bar \nabla \gamma_b C W) V^{b} +  (\bar \psi \gamma^a \nabla_a W)  =0 
\end{eqnarray}
Using the parametrization \eqref{GTAB}, we find  $\bar\nabla_{\alpha} W^{\beta}  = 
(\gamma_{ab})_{\alpha}^{\beta} f^{ab}$, so that $\bar \nabla \gamma_b C W =tr(\gamma_{b} C \gamma_{ac}) f^{ac}=0$ since the trace vanishes. 
We finally obtain the usual free equations of motion
\begin{eqnarray}
\label{gauBB}
\nabla^a f_{ab} = 0\,, ~~~~~ \gamma^a \nabla_a W =0
\end{eqnarray}
Since $\bar\nabla_{\alpha} W^{\beta}  = 
(\gamma_{ab})_{\alpha}^{\beta} f^{ab}$, the field strength $f_{ab}$ is not independent of 
$W$, therefore the first equation (which looks like the Maxwell equations for the superfield $f_{ab}$) is 
a high-derivative equation $\nabla^{b} (\nabla\gamma_{ab}W) = \eta_{ab} \nabla^{b}
(\nabla_{\alpha} W^{\beta}) - \nabla_{\alpha}(\gamma_{a} \gamma^{b} \nabla_{b} W) =0$, but it is satisfied if the Dirac equation is satisfied (the first term vanishes because of the Bianchi identities). 

Therefore, we have 
\begin{eqnarray}
\label{gauC}
F\wedge \star F = (f_{ab} f^{ab} +\lambda^{-1} \bar W C W) V^4 \delta^4(\psi) 
\end{eqnarray}
which can be naturally integrated on the supermanifold 
leading to the high-derivative action
\begin{eqnarray}
\label{gauD}
\int_{\mathcal{SM}^{{(4|4)}}} F\wedge \star F  = \int (f_{ab} f^{ab} +\lambda^{-1} \bar W C W)  = 
\int (\nabla \gamma_{ab} W \nabla \gamma^{ab} W +\lambda^{-1} \bar W C W) 
\end{eqnarray}
After integration on the integral forms we are left with the conventional superspace integral for a function. The action starts with the conventional $f_{ab}f^{ab}$ term.
In order to compute the Berezin integral one has to expand $f_{ab}f^{ab}$ into powers of $\theta$'s, and therefore
the final result will contain higher derivatives of the field strength. The second term however is the conventional superspace action, recovered in the limit $\lambda \rightarrow 0$.
The equations of motion are easily retrieved by introducing the potentials $A_a, A_\alpha$ and 
varying the action with respect to them. 

Let us consider now the $N=2$ case. The field strength (after adopting the conventional 
constraints) is expanded as
\begin{eqnarray}
\label{enne2A}
F = f_{ab} V^a V^b + \overline W^A \gamma_a \psi_A V^a + \Phi^{AB} \bar\psi_A \gamma_5 \psi_B\,.   
\end{eqnarray}
and its Hodge dual is 
\begin{eqnarray}
\label{enne2B}
\star F &=& 
f_{ab} \eta^{aa'} \eta^{bb'} 
\epsilon_{a'b'cd} V^c V^d \delta^8(\psi) +  \lambda^{-1}
(V_3)^a \epsilon_{AA'}  (\overline W^A \gamma_a C\iota^{A'}) \delta^8(\psi) \nonumber \\& -&\lambda^{-2} 
V_4  \epsilon_{A A'}  \epsilon_{B B'} \phi^{AB} \bar\iota^{A'} \gamma_5 \iota^{B'} \delta^8(\psi)
\end{eqnarray}
In order to take into account the correct mass dimensions, we introduce again the scale $\lambda$. The 
action takes the form 
\begin{eqnarray}
\int_{\mathcal{SM}^{{(4|4)}}} F\wedge \star F  = \int (f_{ab} f^{ab} + 
 \lambda^{-1} \bar W^{A} C W_A - \lambda^{-2} \bar \Phi^{AB} \Phi_{AB}) 
 \end{eqnarray}
and is a high-derivative action. 
The same formula cannot be written for higher $N\geq 3$ since 
the mass dimensions exceed those of the integration measure. This is 
consistent with the common lore that there is no superspace action 
for super Yang-Mills with N=4 extended supersymmetry.

\subsection{Supergravity}

Finally, we use the Hodge dual operator also for the construction of the supergravity action.
It has been observed in \cite{GGRS} that 
the full supergravity action with auxiliary fields can be expressed in terms of the superdeterminant. 
To write the supergravity action in our framework we observe that, in the 
case of a curved supermanifold, the volume form can be written as 
\begin{eqnarray}
\label{sdeA}
{\rm Vol}^{(4|4)} = V^{4} \delta^{4}(\psi) = {\rm Sdet}(E) d^{4}x\delta^{4(d\theta)}
\end{eqnarray}
where we express the vielbein $V^{a}$ and the graviton $\psi^{\alpha}$ in terms 
$E^{A}=(V^{a},\psi^{\alpha})$ on a curved basis:
\begin{align}
V^{a}  &  =E_{m}^{a}dx^{m}+E_{\mu}^{a}d\theta^{\mu}%
\nonumber\\
\psi^{\alpha}  &  =E_{m}^{\alpha} dx^{m}+E^{\alpha}_{\mu} d\theta^{\mu}%
\label{SupA}%
\end{align}
where all components are superfields and 
\begin{align}
\mathrm{Sdet}(E) = \frac{\det(E^a_m - E^a_\mu 
(E^{-1})^\mu_\alpha E^\alpha_m)}{\det( E_\mu^\alpha)}
\label{SupB}%
\end{align}

The $(4|4)$-form ${\rm Vol}^{(4|4)}$ is trivially closed (being a top integral form) and, if it represents a cohomology class, it is not exact. Then
\begin{eqnarray}
\label{SupBA}
\int_{\mathcal{SM}^{(4|4)}} \star 1 = \int_{\mathcal{SM}^{(4|4)}} {\rm Vol}^{(4|4)} = \int \mathrm{Sdet}(E)(x, \theta) [d^4xd^4\theta] 
\end{eqnarray}
where the second integration is performed on the superspace coordinates $(x^m, \theta^\mu)$. The symbol 
$[d^4xd^4\theta]$ denotes only on which coordinates the integral has to be performed, 
but it does not represent a measure. Since, as discussed in the literature \cite{GGRS}, the Berezin integral of the superdeterminant in \eqref{SupBA} yields the action of $D=4$ supergravity, this action can be written as 
the integral of the Hodge dual of a constant (Newton's constant). In the next sections, we derive this 
formula from the group-manifold approach. 

In order to make contact with other superspace formulations of supergravity, it is convenient to 
define also the chiral/anti-chiral volume forms (see also \cite{CCG}) as follows 
\begin{eqnarray}
\label{YQA}
{\rm Vol}^{(4|2)}_L &=&  \epsilon_{abcd} V^a\wedge\dots \wedge V^d \delta^2(\psi_L) =  V^4 \delta^2(\psi_L) \nonumber    \,, \\
{\rm Vol}^{(4|2)}_R &=&  \epsilon_{abcd} V^a\wedge\dots \wedge V^d \delta^2(\psi_R)   =  V^4 \delta^2(\psi_R) \,, 
\end{eqnarray}
where $\psi_{L/R} = \frac12 (1\pm \gamma_{5}) \psi$.  
Next, we separate in $V^a = V^a_L + E^a_\mu d\theta^\mu_L$  the part of the vielbein along $d\theta^\alpha_L$ 
(in the flat space, this means $V^a_L = dx^a + \bar\theta_L \gamma^a \psi_R$) and, similarly, we write 
$\psi^\alpha_L =E^\alpha_\mu d\theta^\mu_L + \Lambda^\alpha_a V^a_L$. Notice 
that in general, $\psi^\alpha_L =E^\alpha_\mu d\theta^\mu_L + \Lambda^\alpha_a V^a_L 
+ \Lambda^\alpha_\beta d\theta^\beta_R$ 
but the last term can be set to zero by choosing a suitable gauge (reached by a superdiffeomophism) 
known as chiral/antichiral representation (see \cite{GGRS,Kuzenko:1998tsq}).
Then, using the flat basis $dx^{a}, d\theta_{L}^{\alpha}, d\theta_{R}^{\alpha})$, 
we have 
\begin{eqnarray}
\label{YR}
{\rm Vol}^{(4|2)}_L &=& \epsilon_{abcd} (V^a_L + E^a_\mu d\theta^\mu_L)\wedge\dots \wedge (V^d_L + E^d_\mu d\theta^\mu_L) 
 \delta^2(E^\alpha_\mu d\theta^\mu_L + \Lambda^\alpha_a V^a_L) \\
&=& \epsilon_{abcd} \frac{(V^a_L + E^a_\mu d\theta^\mu_L)\wedge\dots \wedge (V^d_L + E^d_\mu d\theta^\mu_L)}{\det(E^\alpha_\mu)}
 \delta^2(d\theta^\mu_L + (E^{-1})_\alpha^\mu \Lambda^\alpha_a V^a_L) \\
 &=&  \frac{\det(\delta^a_b + E^a_\mu (E^{-1})_\alpha^\mu \Lambda^\alpha_b )}{\det(E^\alpha_\mu)}  
 \epsilon_{abcd} V^a_L\wedge\dots \wedge V^d_L\delta^2(d\theta^\mu_L) \nonumber \\
 &=&  {\cal E}_L \epsilon_{abcd} V^a_L\wedge\dots \wedge V^d_L\delta^2(d\theta^\mu_L) =
  {\cal E}_L V^4_L\delta^2(d\theta^\mu_L) 
 \\\nonumber \\
  {\cal E}_L  &=&  \frac{\det(\delta^a_b + E^a_\mu (E^{-1})_\alpha^\mu \Lambda^\alpha_b )}{\det(E^\alpha_\mu)}
 \end{eqnarray}
where $ {\cal E}_{L}(x, \theta_L)$ is a chiral superfield\footnote{Several textbooks use the chiral coordinates $z_L =(x_L,\theta_L)$ and identify the chiral measure with the Berezinian of the transformation $z'_L = z'_L(z_L)$.} according to our definition. Indeed we have 
\begin{eqnarray}
\label{YRA}
0= d {\rm Vol}^{(4|2)}_L = d {\cal E}_L\wedge V^{4}_{L} \delta^{2}(\psi_{L}) = d \ln {\cal E}_L  \wedge {\rm Vol}^{(4|2)}_L 
\end{eqnarray} 
implying that $ \ln {\cal E}_L$ is a chiral superfield. In the same way, we 
can study ${\rm Vol}^{(4|2)}_{R}$ and the chiral density $ {\cal E}_R$. 

Let us now relate the volume form ${\rm Vol}^{(4|4)}$ to the chiral ones. The superdeterminant 
$E = {\rm Sdet}(E)$ is a function of $(x,\theta,\bar{\theta})$ and reads
\begin{equation}
\mathrm{Sdet}(E)=\frac{\mathrm{det}\Big(E_{m}^{a}-E_{\mu}^{a}(E^{-1})_{\beta
}^{\mu}E_{m}^{\beta}-E_{\dot{\mu}}^{a}(\bar{E}^{-1})_{\dot{\beta}}^{\dot{\mu}%
}E_{m}^{\dot{\beta}}\Big)}{\mathrm{det}(E_L)\mathrm{det}({E}_{R})}=
\frac{\mathrm{Sdet}_{R}(\hat{E})}{\mathrm{det}%
({E}_{R})}=\frac{\mathrm{Sdet}_{L}(\hat{E})}{\mathrm{det}%
({E}_{L})}\label{theoG}%
\end{equation}
where $\mathrm{Sdet}_{L}(\hat{E})$ is the chiral super determinant written in
terms of a redefined vielbein $\hat{E}_{m}^{a}=E_{m}^{a}-E_{R{\mu}}%
^{a}({E}^{-1})_{L {\beta}}^{{\mu}}E_{L m}^{{\beta}}$. In terms of the 
volume form we have
\begin{eqnarray}
{\rm Vol}^{(4|4)} &=& \frac12 V^{4}_{L} \delta^{2}(\psi_{L}) \frac{1}{\mathrm{det}(E_{R})} \delta^{2}(d\theta_{R}) + 
\frac12 V^{4}_{R} \delta^{2}(\psi_{R}) \frac{1}{\mathrm{det}(E_{L})} \delta^{2}(d\theta_{L})
\nonumber \\
&=& \frac12 {\rm Vol}^{(4|2)}_{L} \frac{1}{\mathrm{det}(E_{R})} \delta^{2}(d\theta_{R}) + 
\frac12 {\rm Vol}^{(4|2)}_{R} \frac{1}{\mathrm{det}(E_{L})} \delta^{2}(d\theta_{L})
\label{theoE}%
\end{eqnarray}
where $V_{L}^{a}, \psi_{L}^{\alpha}$ and $V_{R}^{a}, \psi^{\alpha}_{R}$ 
are given in terms of the redefined supervielbeins. 
Then, using the fact that $ \delta^{2}(d\theta_{R})$ and $ \delta^{2}(d\theta_{L})$ 
are not the PCO's, unless they are multiplied by $\theta_{R}^{2}$ and $\theta_{R}^{2}$,
we can expand $1/{\mathrm{det}(E_{R})}$ and  $1/{\mathrm{det}(E_{L})}$ up to second order 
(we recall that $\delta^{2}(d\theta_{R})$ and $\theta_{R}^{\alpha} \delta^{2}(d\theta_{R})$ 
are cohomologically trivial), and since ${\rm Vol}^{(4|2)}_{L}$ and ${\rm Vol}^{(4|2)}_{R}$ 
are closed, we can always discard $d$-exact terms. Therefore, we are left 
with 
\begin{eqnarray}
{\rm Vol}^{(4|4)}&=& {\rm Vol}^{(4|2)}_{L} D^{2}_{R}\left(
\frac{1}{2 \mathrm{det}(E_{R})} \right) \mathbb{Y}_{R} + 
{\rm Vol}^{(4|2)}_{R} D^{2}_{L} \left(\frac{1}{2 \mathrm{det}(E_{L})}\right)  \mathbb{Y}_{L}
\label{theoEA}%
\end{eqnarray}
where $ \mathbb{Y}_{L}$ and $ \mathbb{Y}_{R} $ are the PCO's. 
The two expressions $D^{2}_{R}({1}/2 {\mathrm{det}(E_{R})} )$ and $D^{2}_{L}({1}/2{\mathrm{det}(E_{L})})$ 
are computed in \cite{GGRS,Kuzenko:1998tsq} and are identified with the superfields $\mathcal{R}_{R}$ and 
$\mathcal{R}_{L}$. The superfields $\mathcal{R}_{R/L}$ contain the
auxiliary fields and the Ricci scalar; they appear in the commutation
relations $\{\nabla_{\alpha},\nabla_{\beta}\}=-\bar{\mathcal{R}}\mathcal{M}%
_{\alpha\beta}$, and are components of the torsion $T^{A}$ as will be discussed in the 
forthcoming Section. Finally, we can write the volume form as 
\begin{eqnarray}
{\rm Vol}^{(4|4)}&=& {\rm Vol}^{(4|2)}_{L} \mathcal{R}_{R} \mathbb{Y}_{R} + 
{\rm Vol}^{(4|2)}_{R} \mathcal{R}_{L} \mathbb{Y}_{L} 
\label{theoEB}%
\end{eqnarray}
The formula reproduces, in terms of integral forms, the Siegel chiral-integration formula \cite{GGRS}. 
In superspace language this formula reads 
\begin{eqnarray}
\label{supME}
\int \mathrm{Sdet}(E) = \int_{L} \mathrm{Sdet}_{L}(\hat{E}) {\mathcal{R}_{L}} + 
 \int_{R} \mathrm{Sdet}_{R}(\hat{E}) {\mathcal{R}_{R}}\,, 
\end{eqnarray}
which turns out to be crucial in comparing the geometric formulation of supergravity 
with its superspace formulation (a clear derivation and discussion in the context of conformal supergravity 
can be found in \cite{Kuzenko:1998tsq}). 

\section{Old minimal supergravity}

In supergravity the supervielbeins $E^a, E^\alpha$ become dynamical 
and satisfy equations of motion. In the present Section, we review the 
group manifold formulation of D=4 N=1 supergravity, we define a suitable PCO 
and we finally show the equivalence with the superspace formulation. In the end, we 
write the action in terms of the Hodge dual operator. 

\subsection{Off shell degrees of freedom}

The theory contains a vielbein 1-form $V^a$ with 6 off-shell degrees of freedom 
and a Majorana gravitino $\psi^\al$ with 12 off-shell degrees of freedom.
We can match off-shell d.o.f. by adding 
 three 0-form auxiliary fields: an axial vector $A^a$ with 4 d.o.f., a scalar $S$ with one d.o.f., and 
 a pseudoscalar $P$ with 1 d.o.f.. This set of auxiliary fields was first introduced in \cite{FPvN}. Here we 
reformulate the theory in the group manifold approach. 

\subsection{The algebra and Bianchi identities}
We start from the super Poincar\'e algebra, extended with the three 0-forms $A^a,S,P$.
The deformed Maurer-Cartan equations for this extended superalgebra are given by
\eqa
& & R^{a}= dV^{a} - \omega^{a}_{~c} V^{c} - \frac{i }{2} \bar\psi\gamma^{a} \psi \equiv \Dcal V^a  - \frac{i }{2} \bar\psi\gamma^{a} \psi\label{RasuperPoincare}\\
& & R^{ab}=d \omega^{ab} - \omega^{a}_{~c} ~\omega^{cb} \label{RabsuperPoincare}\\
& & \rho= d \psi- \frac{1}{ 4} \omega^{ab} \gamma_{ab} \psi \equiv \Dcal \psi \label{rhosuperPoincare}\\
& & R(A)^a = dA^a - \omega^{a}_{~c} A^{c} \\
& & R(S) = dS \\
& & R(P) = dP
\ena
These equations can be seen as the definition of curvatures. Taking the exterior derivative of both sides yields the Bianchi identities: 
\eqa
& & dR^a -\omega^a_{~b} R^b + R^a_{~b} V^b - i \psibar \gamma^a \rho \equiv  \Dcal R^a + R^a_{~b} V^b - i \psibar \gamma^a \rho= 0 \label{BianchiRasuperPoincare}\\
& & dR^{ab} - \omega^a_{~c} R^{cb} +  \omega^b_{~c} R^{ca} \equiv \Dcal R^{ab}=0 \label{BianchiRabsuperPoincare}\\
& & d\rho - {1 \over 4} \omega^{ab} \gamma_{ab} \rho + {1 \over 4} R^{ab} \gamma_{ab} \psi \equiv \Dcal \rho 
+ {1 \over 4} R^{ab} \gamma_{ab} \psi =0 \label{BianchirhosuperPoincare} \\
& & \Dcal R(A)^a + R^{ab} A_b =0 \\
& & dR(S)=0 \\
& & dR(P) =0
\ena
The Maurer Cartan equations, and therefore also the Bianchi identities, are invariant under the rescalings
 \eq
     \omega^{ab} \rightarrow \lambda^0 \omega^{ab}, ~V^a \rightarrow \lambda  V^a,~\psi \rightarrow \lambda^{1\over 2} 
     \psi, ~ A^{a}  \rightarrow \lambda^{-1} A^{a}, ~  S  \rightarrow \lambda^{-1} S, ~
     P  \rightarrow \lambda^{-1} P,
     \label{rescalingsold}
      \en          

\subsection{Parametrizations of the curvatures}

 According to the group manifold approach, we parametrize the curvatures so that their outer components
      (i.e. components along at least one fermionic direction) are related to the inner components (i.e. components along bosonic directions) and to the auxiliary fields.  A parametrization compatible with the scalings (\ref{rescalingsold}) and Lorentz invariance is given by:
 \eqa
 & & R^a=0  \label{param1}\\
 & & R^{ab}= R^{ab}_{~~cd} V^c V^d + \thetabar^{ab}_c ~\psi ~V^c  + {c_2 \over 2} \psibar (\gamma^{ab} \eta + \eta \gamma^{ab}) \psi \label{param2}\\
 & & \rho= \rho_{ab}~ V^a V^b  + i c_3 \gamma_5 \psi V^a A_a + i c_4 \gamma_a \eta' \psi V^a \label{param3} \\
 & & R(A)^a = (\Dcal A^a)_b V^b + \psibar \zeta^a  \label{param4}\\
 & & R(S) = (\partial_a S) V^a + \psibar \xi  \label{1c} \label{param5}\\
 & & R(P) = (\partial_a P) V^a + \psibar \chi \label{2c} \label{param6}
 \ena
with real $c_2,c_3,c_4$. The only a priori choice is $R^a=0$, i.e. vanishing (super)torsion.
It can be shown that $R^a=R^a_{bc} V^b V^c$ would only lead to a redefinition of the spin connection,
in terms of $V$ and $\psi$, while outer components of $R^a$ cannot be found using inner components
and auxiliary fields, due to scaling and Lorentz index structure. The Bianchi identities will fix
\eq
\thetabar^{ab}_c  \equiv 2i \rhobar_c^{~[a} \gamma^{b]} - i \rhobar^{ab} \gamma_c,~~~\eta = \eta' \equiv S-i \gamma_5 P + A^a \gamma_a \gamma_5 \label{thetaeta}
 \en
 and $c_2,c_3$ in terms of $c_4$, which remains the only free parameter.
It is convenient to write the gravitino field strength $\rho$ as follows
\begin{eqnarray}
\label{gravFS}
\rho^\alpha \equiv\Dcal\psi^\alpha =  \rho^\alpha_{~ab} V^a V^b + (\rho_{a}\psi)^\alpha V^a
\end{eqnarray}
where $ (\rho_{a}\psi)^\alpha =  \rho^\alpha_{~a \beta} \psi^\beta$ 
and the general form of the matrix $ \rho^\alpha_{~a \beta}$ is read off from 
the parametrization of the curvatures 
$$
\rho^\alpha_{~~a \beta} = i c_4 (\gamma^\alpha_{a \beta} S  - i (\gamma_a \gamma_5)^\alpha_{~\beta} P + (2 (\gamma_5)^\alpha_{~\beta} \delta^b_a - 
(\gamma_5  \gamma^{~b}_a)^\alpha_{~\beta} )A_b )
$$
or, suppressing the spinorial indices 
\begin{eqnarray}
\label{gravFSA}
\rho_{a} &=& i c_4 \Big(\gamma_{a} S  - i \gamma_a \gamma_5 P + \gamma_5 (2 \delta^b_a -   \gamma^b_a) A_b \Big) 
\nonumber \\
&=&  i c_4 \Big(\frac12 \gamma_{a} (1-\gamma_5)(S + i P) + 
 \frac12 \gamma_{a} (1+\gamma_5)(S - i P) + \gamma_5 (2 \delta^b_a -   \gamma^b_a) A_b \Big)
 \nonumber \\
&=& i c_4 \Big(\gamma_a P_L \mathcal{R} + \gamma_a P_R \overline{\mathcal{R}} + \gamma_5 (2 \delta^b_a -   \gamma^{~b}_a) A_b \Big)
\end{eqnarray}
with $P_L , P_R, {\mathcal{R}} ,{\overline {\mathcal{R}}}$ defined by the last equality.
It is a function of the auxiliary fields $S, P, A_a$. Notice that if we give a vacuum expectation value to 
$S$ denoted by $s$, it reduces to 
$$
\rho_{a}(S=s, P=0, A_a=0) =   i c_4 \gamma_{a} s. 
$$

\subsection{The action}

With the usual methods of the group manifold approach \cite{book}, the Lagrangian is  found to be
\begin{equation}
{\mathcal L}^{(4|0)}= R^{ab} V^{c} V^{d} \epsilon_{abcd} + 4
\bar\psi\gamma_{5} \gamma_{a} \rho V^{a}  + \alpha (S^2+P^2 + A^a A_a) \epsi_{abcd} V^a V^b V^c V^d   \label{SGaction1}
\end{equation}

The action is obtained by considering the most general $SO(3,1)$ 
            scalar 4-form, invariant under the rescalings in eq. \eqref{rescalingsold} and then 
            requiring that the equations of motion admit the vanishing curvatures solution
             \eq
             R^{ab} = R^a = \rho = R(A)^a=R(S)=R(P)= 0
             \en
                           The remaining parameter $\alpha$ is fixed by requiring the closure of ${\mathcal L}^{(4|0)}$ , i.e. $d{\mathcal L}^{(4|0)} =0$.
               This yields $\alpha = -2(c_4)^2 $ and ensures off-shell closure of the supersymmetry transformations.
               
There are some remarks to be considered. Even though the auxiliary fields appear in the new term in \eqref{SGaction1}, they 
are also hidden in the Lorentz curvature term and the Rarita-Schwinger term. Then, if we define  auxiliary field-independent curvatures 
$R^{ab}_*, \rho_*$ as 
\begin{eqnarray}
\label{afA}
 &&R^{ab}_*=  R^{ab} - {c_2 \over 2} \psibar (\gamma^{ab} \eta + \eta \gamma^{ab}) \psi \nonumber \\
 && \rho_* = \rho  - i c_3 \gamma_5 \psi V^a A_a - i c_4 \gamma_a \eta \psi V^a\,, 
 \label{param3} 
\end{eqnarray}
 we can rewrite the action as 
 \begin{eqnarray}
{\mathcal L}^{(4|0)} &=& 
\epsilon_{abcd}  R^{ab}_* V^{c} V^{d} + 4
\bar\psi\gamma_{5} \gamma_{a} \rho_* V^{a}  \nonumber \\
&+&  \epsilon_{abcd} {c_4} \Big(  S \psibar \gamma^{ab}\psi  V^{c} V^{d} 
- i P \psibar \gamma^{ab} \gamma^5 \psi  V^{c} V^{d} \Big) + 16 i c_4 A_a \psibar \gamma_{b} \psi  V^{c} V^{d} \nonumber \\
 &-& 2 (c_4)^2 (S^2+P^2 + A^a A_a) \epsilon_{abcd} V^a V^b V^c V^d  \,. 
  \label{SGaction1A}
\end{eqnarray}
In the second and third lines, we have made explicit the auxiliary fields. The first line is the 
usual rheonomic N=1 supergravity action. Notice that the $S$ field could acquire a v.e.v. (which can be achieved also by shifting 
$S \rightarrow S + \frac{1}{2 l}$) and setting $S=A^a=P=0$ we obtain the supergravity action with a cosmological term
 \begin{eqnarray}
\hspace{-.5cm}
{\mathcal L}^{(4|0)}_{AdS} &=& 
\epsilon_{abcd}  R^{ab}_* V^{c} V^{d} + 4
\bar\psi\gamma_{5} \gamma_{a} \rho_* V^{a}  
+\epsilon_{abcd}  \frac{c_4}{2 l} \psibar \gamma^{ab}\psi  V^{c} V^{d} 
- \frac{2 (c_4)^2 }{(2 l)^2} \epsilon_{abcd} V^a V^b V^c V^d   \label{SGaction1B}
\end{eqnarray}
The third term is the usual gravitino mass term needed for supersymmetry 
with the cosmological term.\footnote{By using the relation $\epsilon_{abcd}\gamma^{cd} = 2 i \gamma^5 \gamma_{ab}$, 
and $c_4 = -1/2$, we match the conventional $AdS_4$ expression.}  The resulting action describes N=1 supergravity on $AdS_4$. The positive constant $l$ is the radius of $AdS_{4}$. 
The action turns out to be invariant under the isometry supergroup $OSp(1|4)$, gauge-invariant under local 
Lorentz symmetry and local supersymmetry transformations. The vielbein $V^a$, the gravitino $\psi^\alpha$ and 
the spin connection $\omega^{ab}$ (fixed in terms of $V^a, \psi^\alpha$) are the Maurer-Cartan forms of 
the $\mathfrak{osp}(1|4)$ superalgebra. The spin connection is the gauge connection of the Lorentz group and 
therefore we can consider $V^a, \psi^\alpha$ as the vielbeins of the coset manifold $OSp(1|4)/SO(1,3)$. 
While ${\mathcal L}^{(4|0)}$ in \eqref{SGaction1} is $d$-closed  in presence 
of auxiliary fields (see the forthcoming subsection), the new ${\mathcal L}^{(4|0)}_{AdS}$ is no longer $d$-closed.

\subsection{Fixing coefficients}

From Bianchi identity (\ref{BianchiRasuperPoincare}) for $R^a$, after substituting the parametrizations, one finds in the $\psi VV$ sector the expression for $\thetabar$ given in (\ref{thetaeta}). In the $\psi \psi V$ sector one obtains the relation
\eq
c_2 = c_4
\en
Considering then the Bianchi identity for $\rho$, the $\psi \psi \psi$ sector yields
\eq
c_3 = 3 c_4 
\en
while the $\psi \psi V$ sector fixes $\zeta^a, \xi, \chi$ in the parametrizations:
\eqa
& &  \zeta^a = {1 \over c_4} (  {1 \over 3} \gamma_5 \gamma_b \rho^{ab} - { i \over 12} 
\epsi^{abcd} \gamma_b \rho_{cd} )\\
& & \xi = -{1 \over 6 c_4} \gamma^{ab} \rho_{ab} \\
& & \chi = {i \over 6 c_4}  \gamma_5 \gamma^{ab} \rho_{ab}
 \ena
As a consequence of Bianchi's identities, the relations
\begin{eqnarray}
\label{reA}
\gamma_a \zeta^a - 2 i \chi =0\,,  ~~~~~
\gamma_5 \xi + i \chi =0\,. 
\end{eqnarray}
follow. 
In superspace language \cite{WB} there is the identification 
${\cal R}_{L} = S + i P$ and ${\cal R}_{R} = S - i P$. 
The two superfields ${\cal R}_{L}, {\cal R}_{R}$ are chiral and antichiral, {\it i.e.}
$\nabla_R {\mathcal R}_L = \nabla_L {\mathcal R}_{R} =0$ as a consequence 
of the Bianchi identities. Using the decomposition $\nabla_L =\frac12 (1+ \gamma^5) \nabla$ and 
$\nabla_R =\frac12 (1- \gamma^5) \nabla$, where $\nabla$ is 
the covariant spinorial derivative, we finally find
\begin{eqnarray}
\label{appA}
\nabla S - i \gamma^5 \nabla P =0
\end{eqnarray}
and with the identifications \eqref{param1} we obtain eqs. \eqref{reA}. In addition, 
another set of equations is reported in \cite{GGRS}:
\begin{eqnarray}
\label{appB}
\nabla^\alpha G_{\alpha\dot\alpha} + \overline\nabla_{\dot\alpha} {\mathcal R}_R =0\,, ~~~~~
\overline\nabla^{\dot\alpha} G_{\alpha\dot\alpha} + \nabla_{\alpha}{\cal R}_{L} =0\,, ~~~~~
\end{eqnarray}
with $G_{\alpha\dot\alpha}$ real superfield, 
again as a consequence of the Bianchi identities. Identifying $G_{\alpha\dot \alpha}$ with the 
vector field $A_a \gamma^a_{\alpha \dot\alpha}$ reproduces the first of eqs. \eqref{reA}.

\subsection{Closure of the Lagrangian}

Using the Bianchi identities 
(\ref{BianchiRabsuperPoincare}) and (\ref{BianchirhosuperPoincare}), and the definition 
of the torsion $R^a$ in (\ref{RasuperPoincare}) we find:
\begin{align}
& d{\mathcal L}^{(4|0)}= 2 R^{ab} R^c V^d \epsi_{abcd} + i R^{ab} \psibar \ga^c \psi V^d \epsi_{abcd}+
4 \rhobar \ga_5 \ga_a \rho V^a + \nonumber \\
&  ~~~~~~~ + \psibar \ga_5 \ga_c \ga_{ab} \psi R^{ab} V^c -4 \psibar \ga_5 \ga_a \rho R^a - 2i \psibar \ga_5 \ga_a \rho \psibar \ga^a \psi \nonumber  \\
&   ~~~~~~~ - 4i (c_4)^2 \epsi_{abcd} \psibar \gamma^a \psi V^b V^c V^d (S^2 + P^2 + A^a A_a) \nonumber \\
&   ~~~~~~~-  4 (c_4)^2 (S dS +P dP  + A^a \Dcal A_a) \epsi_{abcd} V^a V^b V^c V^d \label{dL1}
\end{align}
The gamma matrix identity
\eq
\ga_c \ga_{ab} = \eta_{ac} \ga_b - \eta_{bc} \ga_a +i \epsi_{abcd}\ga_5 \ga^d 
\en
implies $\psibar \ga_5 \ga_c \ga_{ab} \psi =i \epsi_{abcd} \psibar \gamma^d \psi$, so that the second and the fourth term cancel in 
(\ref{dL1}). Moreover the Fierz identity
\eq
\gamma_a  \psi \psibar \gamma^a \psi =0 \label{fierz1}
\en
and $\psibar \gamma_5 \gamma_a \rho = \rhobar \gamma_5 \gamma_a \psi$ imply that also the sixth term in
(\ref{dL1}) vanishes. Using then the parametrization $R^a =0$ leads to
\eqa
& & d{\mathcal L}^{(4|0)}= 4 \rhobar \ga_5 \ga_a \rho V^a 
  - 4i (c_4)^2 \epsi_{abcd} \psibar \gamma^a \psi V^b V^c V^d (S^2 + P^2 + A^a A_a) \nonumber  \\
  & & ~~~~ -  4 (c_4)^2 (S dS +P dP  + A^a \Dcal A_a) \epsi_{abcd} V^a V^b V^c V^d  \label{dL2}
\ena
Finally substituting into (\ref{dL2})  the parametrizations for $\rho$ and $dS, dP, \Dcal A_a$ we can check that all  terms cancel, and therefore 
\eq
d{\mathcal L}^{(4|0)} =0
\en
The only remaining free parameter $c_4$ essentially sets the scale of the auxiliary fields
(changing its value amounts to rescale $\eta$) and can be chosen as
\eq
c_4 = {1 \over 6}
\en
to make contact with the notations of ref.\cite{FPvN}.

\subsection{Picture Changing Operators}

To compute the superspace action starting from the rheonomic Lagrangian \eqref{SGaction1}, we have to introduce a new Picture Changing Operator. From the analysis performed in \cite{CCG}, we know that in $D=4, N=1$ the non-trivial PCO must have the form:
\begin{equation}
	\mathbb{Y}^{(0|4)} \sim \theta^2 V^2_0 \iota_{D}^2 \delta^4 \left( \psi_0 \right) \ ,
\end{equation}
with an explicit dependence on the $\theta$ coordinates. We used the flat supervielbeins 
\begin{eqnarray}
\label{cicA}
V^a_0 = dx^a + \bar\theta \gamma^a d\theta\,, ~~~~~~~~
\psi^{\alpha}_0 = d \theta^{\alpha}\,. 
\end{eqnarray}
Notice that the form number carried by the vielbeins $V_0^a$ is compensated by the negative form number carried by the contractions along odd vector fields $\iota_{D}$. This structure turns out to be the right expression to translate the group manifold actions for WZ and SYM $N=1,2$ into the corresponding well-known superspace actions \cite{CCG}. 
Notice that although the explicit dependence on the $\theta$'s might indicate a supersymmetry breaking,  it turns out that it corresponds to the chiral/anti-chiral projections to sub-superspaces which are supersymmetric invariant. The PCO can be built out of the following terms:

\begin{eqnarray}\label{YA}
	\mathbb{Y}_S^{(0|4)} &=& \bar{\theta} \theta V_0^a \wedge V_0^b \bar{\iota} \gamma_{ab} \iota \delta^4 \left( \psi_0\right) \ ,\\
	 \mathbb{Y}_P^{(0|4)} &=& \bar{\theta} \gamma_5 \theta V_0^a \wedge V_0^b \bar{\iota} \gamma_{ab} \gamma_5 \iota \delta^4 \left( \psi_0\right) \ , \\
	 \mathbb{Y}_A^{(0|4)} &=& \epsilon_{abcd} \bar{\theta} \gamma^a \gamma_5 \theta V_0^b \wedge V_0^c \bar{\iota} \gamma^d \iota \delta^4 \left( \psi_0\right) \ . \\
	 \mathbb{Y}_{\hat{S}}^{(0|4)} &=& \bar{\theta} \theta  \epsilon_{abcd} V_0^a \wedge V_0^b \bar{\iota} \gamma^{cd} \iota \delta^4 \left( \psi_0\right) \ ,\\
	 \mathbb{Y}_{\hat{P}}^{(0|4)} &=& \bar{\theta} \gamma_5 \theta \epsilon_{abcd} V_0^a \wedge V_0^b \bar{\iota} \gamma^{cd} \gamma_5 \iota \delta^4 \left( \psi_0\right) \ , \\
	 \mathbb{Y}_{\hat{A}}^{(0|4)} &=& \bar{\theta} \gamma_a \gamma_5 \theta V_0^a \wedge V_0^b \bar{\iota} \gamma_b \iota \delta^4 \left( \psi_0\right) \ .
\end{eqnarray}

Let us check which combination is closed. We compute their exterior derivatives:
\begin{eqnarray}\label{YB}
	d \mathbb{Y}_S^{(0|4)} &=& 2 \bar{\psi} \theta V_0^a \wedge V_0^b \bar{\iota} \gamma_{ab} \iota \delta^4 \left( \psi_0\right) + i \bar{\theta} \theta \bar{\psi} \gamma^a \psi_0\wedge V_0^b \bar{\iota} \gamma_{ab} \iota \delta^4 \left( \psi_0\right) \nonumber  \\
	\nonumber  &=& -4 V_0^a \wedge V_0^b  \bar{\theta} \gamma_{ab} \iota \delta^4 \left( \psi_0\right) + 2i \bar{\theta} \theta \text{tr} \left( \gamma^a \gamma_{ab} \right) \delta^4 \left( \psi_0\right) = -4 V_0^a \wedge V_0^b  \bar{\theta} \gamma_{ab} \iota \delta^4 \left( \psi_0\right) \nonumber \\
	d \mathbb{Y}_P^{(0|4)} &=& 2 \bar{\psi} \gamma_5 \theta V_0^a \wedge V_0^b \bar{\iota} \gamma_{ab} \gamma_5 \iota \delta^4 \left( \psi_0\right) + i \bar{\theta} \gamma_5 \theta \bar{\psi} \gamma^a \psi_0\wedge V_0^b \bar{\iota} \gamma_{ab} \gamma_5 \iota \delta^4 \left( \psi_0\right)   \\
	\nonumber  &=& -4 V_0^a \wedge V_0^b  \bar{\theta} \gamma_{ab} \iota \delta^4 \left( \psi_0\right) + 2i \bar{\theta} \gamma_5 \theta \text{tr} \left( \gamma^a \gamma_{ab} \gamma_5 \right) \delta^4 \left( \psi_0\right) = -4 V_0^a \wedge V_0^b  \bar{\theta} \gamma_{ab} \iota \delta^4 \left( \psi_0\right)  \\
	d \mathbb{Y}_A^{(0|4)} &=& 
	2 V_0^a \wedge V_0^b  \bar{\theta} \gamma_{ab} \iota \delta^4 \left( \psi_0\right) + 2i \epsilon_{abcd} 
	\bar{\theta} \gamma^{ab}\gamma_5 \theta \text{tr} \left( \gamma^c \gamma^{d}\right) \delta^4 \left( \psi_0\right)  \nonumber \\&=& 
	2 V_0^a \wedge V_0^b  \bar{\theta} \gamma_{ab} \iota \delta^4 \left( \psi_0\right)  \nonumber \\
d \mathbb{Y}_{\hat S}^{(0|4)} &=& 2 \bar{\psi} \theta \epsilon_{abcd} V_0^a \wedge V_0^b \bar{\iota} \gamma^{cd} \iota \delta^4 \left( \psi_0\right) 
+ i \bar{\theta} \theta  \epsilon_{abcd} \bar{\psi} \gamma^a \psi_0\wedge V_0^b \bar{\iota} \gamma^{cd} \iota \delta^4 \left( \psi_0\right) \nonumber \\&=& 
-4 \epsilon_{abcd} V_0^a \wedge V_0^b  \bar{\theta} \gamma^{cd} \iota \delta^4 \left( \psi_0\right) \nonumber \\
d \mathbb{Y}_{\hat P}^{(0|4)} &=& 2 \bar{\psi} \gamma_5 \theta \epsilon_{abcd} V_0^a \wedge V_0^b \bar{\iota} \gamma^{cd} \gamma_5\iota \delta^4 \left( \psi_0\right) 
+ i \bar{\theta} \gamma_5 \theta  \epsilon_{abcd} \bar{\psi} \gamma^a \psi_0\wedge V_0^b \bar{\iota} \gamma^{cd} \gamma_5\iota \delta^4 \left( \psi_0\right) \nonumber \\&=& 
-4 \epsilon_{abcd} V_0^a \wedge V_0^b  \bar{\theta} \gamma^{cd} \iota \delta^4 \left( \psi_0\right) \nonumber \\
d \mathbb{Y}_{\hat A}^{(0|4)} &=& 
	2 V_0^a \wedge V_0^b  \bar{\theta} \gamma_{ab} \gamma_5 \iota \delta^4 \left( \psi_0\right) + 2i  
	\bar{\theta} \gamma_{a}\gamma_5 \theta V_0^a \delta^4 \left( \psi_0\right) 
	 \nonumber \\&=& 
		2 V_0^a \wedge V_0^b \epsilon_{abcd}   \bar{\theta} \gamma^{cd} \iota \delta^4 \left( \psi_0\right) + 2i  
	\bar{\theta} \gamma_{a}\gamma_5 \theta V_0^a \delta^4 \left( \psi_0\right) 
	\end{eqnarray}
and find that there are three independent closed $\mathbb{Y}^{(0|4)}$'s. 
Namely the most general solution of the equation $d \mathbb{Y}^{(0|4)} =0$ is 
\begin{eqnarray}
\label{YC}
\mathbb{Y}^{(0|4)} = \alpha \mathbb{Y}_S^{(0|4)} + \beta \mathbb{Y}_P^{(0|4)} + 2(\alpha+\beta) \mathbb{Y}_A^{(0|4)} + 
\sigma (\mathbb{Y}_{\hat S}^{(0|4)} - \mathbb{Y}_{\hat P}^{(0|4)})  
\end{eqnarray}
where $\alpha, \beta, \sigma$ are arbitrary constants. However, to compute the cohomology we have to 
check whether \eqref{YC} is exact. We observe that there 
are two possible candidates $\eta^{(-1|4)}_S , \eta^{(-1|4)}_P$,   
\begin{eqnarray}
\label{YD}
\eta^{(-1|4)}_S &=& V_0^a\wedge V_0^b (\bar\theta \theta \bar\theta\gamma_5\iota + \bar\theta\gamma_5\theta \bar\theta\iota) \bar\iota\gamma_{ab}\iota \delta^4(\psi_0) \nonumber \\
\eta^{(-1|4)}_P &=& V_0^a\wedge V_0^b (\bar\theta \theta \bar\theta\gamma_5\iota + \bar\theta\gamma_5\theta \bar\theta\iota) \bar\iota\gamma_{ab} \gamma_5\iota \delta^4(\psi_0)
\end{eqnarray}
whose differentials can be added to  $\mathbb{Y}^{(0|4)}$ as 
$\mathbb{Y}^{(0|4)} + \tau d\eta^{(-1|4)}_S  + \rho d\eta^{(-1|4)}_P$. 
This implies that there is only one cohomology class of 
$H^{(0|4)}(d)$ and 
we can choose the representative of the class by choosing $\tau$ and $\rho$. 
We find it convenient to adopt the following representative 
\begin{eqnarray}
\label{YE}
\mathbb{Y}^{(0|4)} &=& \mathbb{Y}_S^{(0|4)} + \mathbb{Y}_P^{(0|4)} + 4 \mathbb{Y}_A^{(0|4)}  \\&=& 
\left(  \bar{\theta} \theta V_0^a \wedge V_0^b \bar{\iota} \gamma_{ab} \iota 
 + \bar{\theta} \gamma_5 \theta V_0^a \wedge V_0^b \bar{\iota} \gamma_{ab} \gamma_5 \iota +  
 \epsilon_{abcd} \bar{\theta} \gamma^a \gamma_5 \theta V_0^b \wedge V_0^c \bar{\iota} \gamma^d \iota\right) \delta^4 \left( \psi_0\right) \nonumber 
\end{eqnarray}
Note that we can combine the first two terms $\mathbb{Y}_S^{(0|4)} + \mathbb{Y}_P^{(0|4)}$ into a chiral and anti-chiral expression. 
Another convenient choice is $\alpha = - \beta$ and $\sigma =0$. This leads to 
\begin{eqnarray}
\label{YEA}
\mathbb{Y}^{(0|4)} &=& \mathbb{Y}_S^{(0|4)} - \mathbb{Y}_P^{(0|4)}  \\&=& 
\left(  \bar{\theta} \theta V_0^a \wedge V_0^b \bar{\iota} \gamma_{ab} \iota 
 - \bar{\theta} \gamma_5 \theta V_0^a \wedge V_0^b \bar{\iota} \gamma_{ab} \gamma_5 \iota\right) \delta^4 \left( \psi_0\right) \nonumber 
\end{eqnarray}
Finally,  note that all representatives are related also to 
\begin{eqnarray}
\label{YF}
\mathbb{Y}^{(0|4)}_{comp} = \bar\theta \theta \bar\theta \gamma_5 \theta \delta^4(\psi_0)  
\end{eqnarray}
which is closed and not exact, but  completely breaks supersymmetry and,  
inserted into the action, yields the component action.  

One may wonder whether it may be possible to build a PCO $\mathbb{Y}^{(0|4)}$ which 
is manifestly supersymmetric invariant, namely written in terms of $V_0^a, \psi^\alpha_{0}$ and 
$\delta^4(\psi_{0})$ and contractions along the $\psi^\alpha_{0}$'s of the Dirac deltas. 
The requirement to be a zero form restricts the possible choices to
\begin{eqnarray}
\label{YG}
\delta^4(\psi_{0})\,, ~~~~
V_0^a\wedge V_0^b \bar\iota \gamma_{ab} \iota \delta^4(\psi_{0})\,, ~~~~
V_0^a\wedge V_0^b \bar\iota \gamma_{ab} \gamma_5\iota \delta^4(\psi_{0})\,.  ~~~~
\end{eqnarray}
These three terms turn out to be exact. Therefore, in the present case, there is no manifestly supersymmetric PCO. 
The same situation occurs also for extended supersymmetric models. 

We want now to extend flat PCO (constructed out of flat $V^{a}_{0}$ and  $\psi^{\alpha}_{0}$) 
to a PCO for a dynamical curved manifold. 
We can deform \eqref{YE} it without changing its cohomology class by acting with a super-diffeomorphism. 
This can be done by acting infinitesimally with a Lie derivative ${\mathcal L}_X$ along a vector field $X$ since 
\begin{eqnarray}
\label{YFA}
\delta \mathbb{Y}^{(0|4)} = {\mathcal L}_X \mathbb{Y}^{(0|4)} = d \left( \iota_X \mathbb{Y}^{(0|4)} \right) 
\end{eqnarray}
or, in the same way, by performing a finite transformation generated by $X$ such that 
\begin{eqnarray}
\label{YFB}
e^{{\mathcal L}_X} V^a_0 = V^a\,, ~~~~~~e^{{\mathcal L}_X} \psi^\alpha_0 = \psi^\alpha\,,~~~~~
e^{{\mathcal L}_X} \theta^\alpha = \Theta^\alpha 
\end{eqnarray}
where $\Theta^\alpha$ is the curved fermionic coordinate. Then we have 
\begin{eqnarray}
\label{YFC}
\mathbb{Y}^{(0|4)} &\rightarrow&  e^{{\mathcal L}_X} \mathbb{Y}^{(0|4)}  \\
&=& 
\left(  \bar{\Theta} \Theta V^a \wedge V^b \bar{\iota} \gamma_{ab} \iota 
 + \bar{\Theta} \gamma_5 \Theta V^a \wedge V^b \bar{\iota} \gamma_{ab} \gamma_5 \iota +  
 \epsilon_{abcd} \bar{\Theta} \gamma^a \gamma_5 \Theta V^b \wedge V^c \bar{\iota} \gamma^d \iota\right) \delta^4 \left( \psi\right) \nonumber
\end{eqnarray}
or, for the second choice \eqref{YEA}, 
\begin{eqnarray}
\label{YFD}
\mathbb{Y}^{(0|4)} &\rightarrow&  e^{{\mathcal L}_X} \mathbb{Y}^{(0|4)} \\
&=&
\left(  \bar{\Theta} \Theta V^a \wedge V^b \bar{\iota} \gamma_{ab} \iota 
 - \bar{\Theta} \gamma_5 \Theta V^a \wedge V^b \bar{\iota} \gamma_{ab} \gamma_5 \iota\right) \delta^4 \left( \psi\right) \nonumber 
\end{eqnarray}
where $V^a, \psi^\alpha$ are the dynamical vielbeins. These curved  $\mathbb{Y}^{(0|4)}$ belong to the same cohomology class of the original $\mathbb{Y}^{(0|4)}$.\footnote{
The topology of the manifold is unchanged by local diffeomorphisms, and 
therefore we cannot study in this way the case of curved rigid supermanifolds not connected by infinitesimal diffeomorphisms to the flat space.} 
 
\subsection{Superspace Action}

Having found the group manifold Lagrangian using geometric means, we use now the 
PCO's to construct the action. On one side, using the flat non-supersymmetric PCO \eqref{YF}, namely 
$\mathbb{Y}^{(0|4)} = \bar \theta \theta \bar \theta\gamma^5 \theta \delta^4(\psi)$, the action reduces to the 
component supergravity action with auxiliary fields on spacetime. 
On the other side, we would like to use the supersymmetric PCO \eqref{YFD} to obtain 
the superspace action as given in \cite{WB,GGRS}, which is manifestly invariant under local supersymmetry.  For this purpose, 
we use \eqref{YFD} to project the action to only a few terms and then we use the
parametrizations \eqref{param1}-\eqref{param6} to simplify the result and discover that the complete superspace 
action is encoded in the auxiliary superfields terms of \eqref{SGaction1}. 
This allows the comparison with the result given in \cite{WB}. Finally, using the relation between chiral volume forms and the superdeterminant (discussed in sec. 2.4), we show the equivalence with the superspace formulation of \cite{GGRS}. 

By inserting $\mathbb{Y}^{(0|4)}$ into the supermanifold integral, we have
\begin{eqnarray}
\label{YH}
S = \int_{\mathcal{SM}^{(4|4)}} {\mathcal L}^{(4|0)} \wedge \mathbb{Y}^{(0|4)}
\end{eqnarray}
The action $ {\mathcal L}^{(4|0)}$ is a superform and therefore can be expanded in powers of 
$V$ and $\psi$ as follows 
\begin{eqnarray}
\label{YI}
 {\mathcal L}^{(4|0)} &=& {\mathcal L} \epsilon_{abcd} V^a V^b V^c V^d + 
  {\mathcal L}_{\alpha,abc}  \psi^\alpha V^a V^b V^c
+  {\mathcal L}^{I}  (\bar\psi {\mathcal M}_{I,ab}\psi) V^a V^b 
\nonumber \\
&+& {\mathcal L}^I_{\alpha abc}  \psi^\alpha  (\bar\psi {\mathcal M}^{ab}_I\psi)  V^c +
 {\mathcal L}^{IJ}_{ab,cd}  (\bar\psi {\mathcal M}^{ab}_I\psi)(\bar\psi {\mathcal M}^{cd}_J\psi) 
\end{eqnarray}
The indices $I,J$ denote the different Dirac matrix structures for bilinears $(\bar\psi {\mathcal M}^{ab}_I\psi)$.
Inserting these expressions in (\ref{YH})  and using the PCO \eqref{YFD} we are left with 
 \begin{eqnarray}
\label{YN}
S &&\int_{\mathcal{SM}^{(4|4)}} {\mathcal L}^{(4|0)} \wedge \mathbb{Y}^{(0|4)}  \nonumber  \\ 
&&= \int_{\mathcal{SM}^{(4|4)}} \left( {\mathcal L}^{I}  (\bar\psi {\mathcal M}_{I,ab}\psi) V^a V^b 
\right) \wedge \mathbb{Y}^{(4|0)} \\
&& = \int_{\mathcal{SM}^{(4|4)}} \!\!\left(   - S \bar\Theta \Theta 
+i P \bar\Theta \gamma_5 \Theta 
\right) V^4 \delta^4(\psi) \nonumber 
\end{eqnarray}
A crucial role 
is played by the derivative of Dirac delta functions of $\psi$'s in the PCO: 
with integration by parts, only the piece \eqref{YN} is selected. 
To compare the result with superspace literature \cite{WB,GGRS,Kuzenko:1998tsq}
we use the chiral notation $\Theta_{L/R} = \frac12 (1 \pm \gamma_5) \Theta$, and 
rewrite the above expression as 
\begin{eqnarray}
\label{YP}
S = \int_{\mathcal{SM}^{(4|4)}} \!\! \left(   {\mathcal R}_R  \bar\Theta_L \Theta_L  
+  {\cal R}_{L }\bar \Theta_R \Theta_R  
\right) V^4 \delta^4(\psi)
\end{eqnarray}
where $ {\mathcal R}_{L},  {\mathcal R}_R$ are the auxiliary fields as discussed in sec. 3.5, equation 
\eqref{reA}. In \cite{WB,GGRS} a complete description of ${\mathcal R}_{L},  {\mathcal R}_R$ and their 
component expansion is given.
We notice that, using the off-shell parametrization given in \eqref{param1}-\eqref{param6}, the complete action 
\eqref{YI} reduces to the terms containing the auxiliary superfields $S, P$, or equivalently to the ${\mathcal R}_{L},  {\mathcal R}_R$ terms. 
This is a well-known phenomenon (see \cite{CCG}) and the full superspace action is obtained by the 
superspace expansion of the auxiliary field terms. 

The expression \eqref{YP} 
can be conveniently rewritten as 
\begin{eqnarray}
\label{YP}
S = \int_{\mathcal{SM}^{(4|4)}} \!\! \left(  {\mathcal R}_R  {\rm Vol}^{(4|2)}_{R} \wedge \mathbb{Y}_{L} +  
 {\mathcal R}_L {\rm Vol}^{(4|2)}_{L} \wedge \mathbb{Y}_{R}  \right) 
\end{eqnarray}
where we used the notation $ \mathbb{Y}_{R}  = \Theta_R \Theta_R  \delta^{2}(\psi_{R})$ 
and $ \mathbb{Y}_{L}  = \Theta_L \Theta_L  \delta^{2}(\psi_{L})$ to denote the PCO 
projecting on  the constraints $\Theta_R=0$ and $ \Theta_L=0$. In the above equation, we 
used $ {\rm Vol}^{(4|2)}_{R} $ and $ {\rm Vol}^{(4|2)}_{L}$ to denote the chiral densities.  

Finally, integrating on the ``cotangent'' coordinates $(V^{a}, \psi_{L}^{\alpha}, \psi_{R}^{\alpha})$, 
we arrive at the simplified expression \cite{WB}
\begin{eqnarray}
\label{YPA}
S &=&  \int_{L}   {\mathcal R}_L {\cal E}_{L}
+ \int_{R} {\mathcal R}_R {\cal E}_{R}  =  \int E = \int_{\mathcal{SM}^{(4|4)}} \star 1
\end{eqnarray}
where we used a chiral integration formula by W. Siegel \cite{GGRS,Kuzenko:1998tsq} 
giving a relation between the chiral measures $ {\cal E}_{L}, {\cal E}_{R}$  and the superdeterminant $E$. 
The integrals in the above formula: $ \int_{L/R}$ are over the chiral superspaces and  
the integral $\int E$ is extended to the entire superspace. The last expression finally shows the relation  
of the superspace action with the Hodge dual of a constant. Reinstalling the appropriate dimensions, 
this constant coincides with Newton's constant.

\section{New minimal supergravity}

 \subsection{Off shell degrees of freedom}

We can match off-shell d.o.f. by adding an auxiliary bosonic 1-form $A$ (3 d.o.f.) and an auxiliary bosonic 2-form $T$ (3 d.o.f.).  The theory with these auxiliary fields was first constructed in ref. \cite{SW}, and recast
in the group manifold formulation in ref. \cite{DFTvN}.

  \subsection{The extended superPoincar\'e algebra}
  
 The starting superalgebra is the superPoincar\'e algebra, extended with a 1-form $A$ and a 2-form $T$.
 The deformed Cartan-Maurer equations for the extended
  soft superPoincar\'e manifold are
   \eqa
   & & R^{ab}=d \omega^{ab} - \omega^a_{~c} ~ \omega^{cb} \\
   & & R^a=dV^a - \omega^a_{~b} ~ V^b -{i \over 2} \psibar \gamma^a \psi \equiv \Dcal V^a -{i \over 2} \psibar \gamma^a \psi\\
   & & \rho = d\psi - {1 \over 4} \omega^{ab} \gamma_{ab} \psi  - {i \over 2} \ga_5 \psi A \equiv \Dcal \psi 
   - {i \over 2} \ga_5 \psi A\\
   & & R^{\square} = dA \\
   & & R^{\otimes}=dT-{i \over 2} \psibar \gamma_a \psi ~V^a
    \ena
    \noi where $\Dcal$ is the Lorentz covariant derivative. These equations can be considered  {\it definitions} for the Lorentz curvature, the (super)torsion, the gravitino field strength and the 1-form and 2-form field strengths respectively.
    The Cartan-Maurer equations are invariant under rescalings 
     \eq
     \omega^{ab} \rightarrow \lambda^0 \omega^{ab}, ~V^a \rightarrow \lambda  V^a,~\psi \rightarrow \lambda^{1\over 2} \psi,~A \rightarrow \lambda^0 A,~T \rightarrow \lambda^2 T \label{rescalings}
      \en           
    \noi Taking exterior derivatives of both sides yields the Bianchi identities:
     \eqa
    & &  \Dcal R^{ab} =0 \\
    & &  \Dcal R^a + R^a_{~b} ~ V^b - i~ \psibar \gamma^a \rho =0\\
    & & \Dcal \rho + {1 \over 2} \gamma_5 \rho A+ {1 \over 4} R^{ab} \gamma_{ab} ~\psi - {i \over 2} \gamma_5 \psi R^\square=0\\
    & & dR^\square=0 \\
    & & dR^{\otimes} - i~ \psibar \gamma_a \rho V^a + {i \over 2} \psibar \gamma_a \psi~ R^a = 0
     \ena
     invariant under the rescalings (\ref{rescalings}).
     
     \subsection{Curvature parametrizations}
     
     According to the group manifold approach, we again parametrize the curvatures so that ``outer'' components
      (i.e. components along at least one fermionic direction) are related to inner components (i.e. components on bosonic directions).
      The most general parametrization compatible with the scalings (\ref{rescalings}) and $SO(3,1) \times U(1)$ gauge invariance 
   is the following:
       \eqa\label{leaA}
        & & R^{ab} = R^{ab}_{~~cd} ~V^c V^d + \thetabar^{ab}_{~~c}~\psi ~V^c + i  c_1~\epsilon^{abcd} ~\psibar \gamma_c \psi f_d\\
        & & R^a = 0 \\
        & & \rho = \rho_{ab} V^a V^b + i a \gamma_5 \psi f_a V^a - i c_2  \gamma_5 \gamma_{ab} \psi V^a f^b  \\
        & & R^\square = F_{ab} V^a V^b + \psibar \chi_a V^a + i c_3 \psibar \gamma_a \psi f^a \\
        & & R^{\otimes} = f^a~V^b V^c V^d \epsilon_{abcd} \\
        & & \Dcal f_a= (\Dcal_b f_a)~ V^b + \psibar \Xi_a
        \ena
                The $VV$ component $F_{ab}$ of $F$, and the $VVV$ component $f_a$ of $R^{\otimes}$ scale respectively as $F_{ab} \rightarrow \lambda^{-2} F_{ab}$ and $f_a \rightarrow \lambda^{-1} f_a$.  The Bianchi identities require that: 
         \eq
         c_1=  c_2 = {3 \over 2} ,~c_3 = 3-a
         \en
         \noi and
           \eqa
         & & \thetabar^{ab}_{~~c} = 2 i \rhobar_c^{~[a} \gamma^{b]}  -i ~\rhobar^{ab} \gamma_c\\
         & & \Xi^a= - {i \over 3!} \epsilon^{abcd} \gamma_b \rho_{cd} \\
         & & \chi_a= 2 ( \gamma_5 \gamma^b  \rho_{ab} + {ia \over 3!} \epsilon_{abcd} \gamma^b \rho^{cd})
         \ena

         \noi Note that thanks to the presence of the auxiliary fields, the Bianchi identities do not imply 
         equations of motion for the spacetime components of the curvatures.
         
         \subsection{The group manifold action}
         
         With the usual group manifold methods, the action is determined to be
          \eq
           S_{d=4 SG} = \int_{M^4} R^{ab} V^c V^d \epsilon_{abcd} + 4 \psibar \gamma_5 \gamma_a  \rho V^a - 4 R^\square T +
            \alpha (f_a R^{\otimes} V^a +{1 \over 8} f_e f^e  V^a V^b V^c V^d \epsilon_{abcd})
           \label{spacetimeaction}
            \en
            This action is obtained by taking for the Lagrangian ${\mathcal L}^{(4|0)}$  the most general $SO(3,1) \times U(1)$ 
            scalar 4-form, invariant under the rescalings discussed above, and then 
            requiring that the variational equations admit the vanishing curvatures solution
             \eq
             R^{ab} = R^a = \rho = R^\square=R^{\otimes} = f_a= 0
             \en
                           The remaining parameter $\alpha$ is fixed by requiring the closure of ${\mathcal L}^{(4|0)}$ , i.e. $d{\mathcal L}^{(4|0)} =0$.
               This yields $\alpha = 4(4a-3)$ and ensures off-shell closure of the supersymmetry transformations
               given below. Notice that $a$ is essentially free, since the term $ i a \gamma_5 \psi f_a V^a$ in the 
               parametrization of the gravitino curvature $\rho$ can be reabsorbed into the definition of the 
               $SO(3,1) \times U(1)$-covariant derivative on $\psi$, by redefining $A'=A + 2 a f_a V^a$. 
               Choosing $a={3 \over 4}$ simplifies the action, reducing it to the first three terms, so that
               the 0-forms $f_a$ do not appear. 
               
               \subsection{Field equations}
               
               Varying $\omega^{ab}$, $V^a$, $\psi$, $A$, $T$ and $f_{a}$ in the action (\ref{spacetimeaction}) leads to the   
                equations of motion:
                \eqa
                 & & 2 \epsilon_{abcd} R^c V^d =0 ~~\Rightarrow ~R^a=0 \\
                 & & 2 R^{bc} V^d \epsilon_{abcd}-4 \psibar \gamma_5 \gamma_a \rho + \alpha 
                 (-f_a R^\otimes + {1\over 2} f_e f^e \epsilon_{abcd} V^b V^c V^d) = 0 \\
                 & & 8 \gamma_5 \gamma_a \rho V^a - 4 \gamma_5 \gamma_a \psi R^a - i \alpha \gamma_a \psi V^a
                 f_b V^b =0 \\
                 & & R^\otimes = 0 \\
                 & & -4 R^\square + \alpha (V^a \Dcal f_a - {i \over 2} f_a \psibar \gamma^a \psi - f_a R^a) =0\\
                 & & R^\otimes = f^a V^b V^c V^d \epsilon_{abcd}
                 \ena
                 These equations are satisfied by the curvatures parametrized as in Section 9.3
          and also imply:
                  \eq
                  R^a=R^\square = R^\otimes = f_a = 0
                   \en
                   \eqa
                   & & R^{ac}_{~~bc} - {1 \over 2} \delta^a_b R^{cd}_{~~cd} =0 ~~~(Einstein~eq.)\label{EIN1}\\
                   & & \gamma^a \rho_{ab}=0~~~(Rarita-Schwinger~eq.)\label{EIN2}
                   \ena
                   The theory has therefore the same dynamical content as the usual $N=1$, $d=4$ supergravity
                   without auxiliary fields.

               \subsection{Off-shell supersymmetry transformations}
               
               Supersymmetry transformations are obtained by applying the Lie derivative along 
               the fermionic directions (i.e. along tangent vectors dual to $\psi$):
                \eqa
              & &  \delta_\epsi V^a = -i \psibar \gamma^a \epsi \\
              & &  \delta_\epsi  \psi = \Dcal \epsi + {i \over 2} \gamma_5 A \epsi + i a \gamma_5 \epsi f_a V^a - 
               {3 i \over 2} \gamma_5 \gamma_{ab} \epsi V^a f^b\\
               & & \delta_\epsi A= {\bar \epsi } ({ia\over 3} \epsilon^{abcd} \gamma_b \rho_{cd} - 2 \gamma_5 \gamma_b \rho^{ba}) + 2i (3-a) {\bar \epsi } \gamma_a \psi f^a
               V_a \\
             & & \delta \omega^{ab}= \thetabar^{ab}_{~~c} ~ \epsi V^c - 3i \epsilon^{abcd} \psibar \gamma_c \epsi f_d \\
              & & \delta_\epsi  T =  - i \psibar \gamma_a \epsi V^a \\
              & & \delta_\epsi f^a = \epsilonbar ~ \Xi^a
                \ena
               and close on all the fields without the need of imposing the field equations \eqref{EIN1}-\eqref{EIN2}.

\subsection{Superspace Action}

Having found the rheonomic Lagrangian for the new minimal set of auxiliary fields, by using the geometric means of the group manifold approach, we would like again to use the 
PCO to define its variational principle and write the corresponding action. On one side, it 
is easy to check that using the flat non-supersymmetric PCO $\mathbb{Y}^{(0|4)} = \bar \theta \theta \bar \theta\gamma^5 \theta \delta^4(\psi)$, we obtain the usual component action with auxiliary fields. On another side, 
we would like to verify that using the supersymmetric PCO \eqref{YFD} leads to a superspace action for the new minimal 
$D=4$ supergravity. 
The latter has been discussed in \cite{Girardi:1984eq,Muller:1985vga,Muller:1988ux} and appears to be of a 
BF-type action. Therefore, the plan of the present Section is to insert the curved PCO which selects some pieces of the 
Lagrangian and, using the parametrizations, to compare the result with the superspace action given in \cite{Muller:1988ux}.

We begin by inserting $\mathbb{Y}^{(0|4)}$ into the supermanifold integral, and
consider the Lagrangian with $\alpha =0$ for simplicity: 
\begin{eqnarray}
\label{YHnew}
S = \int_{\mathcal{SM}^{(4|4)}} {\mathcal L}^{(4|0)} \wedge \mathbb{Y}^{(0|4)}
\end{eqnarray}
The action $ {\mathcal L}^{(4|0)}$ is a superform and therefore can be expanded in powers of 
$V,\psi$ as follows 
\begin{eqnarray}
\label{YInew}
 {\mathcal L}^{(4|0)} &=& {\mathcal L} \epsilon_{abcd} V^a V^b V^c V^d + 
  {\mathcal L}_{\alpha,abc}  \psi^\alpha V^a V^b V^c
+  {\mathcal L}^{I}  (\bar\psi {\mathcal M}_{I,ab}\psi) V^a V^b 
\nonumber \\
&+& {\mathcal L}^I_{\alpha abc}  \psi^\alpha  (\bar\psi {\mathcal M}^{ab}_I\psi)  V^c +
 {\mathcal L}^{IJ}_{ab,cd}  (\bar\psi {\mathcal M}^{ab}_I\psi)(\bar\psi {\mathcal M}^{cd}_J\psi) 
\end{eqnarray}
The indices $I,J$ denote the different Dirac matrix structures for bilinears $(\bar\psi {\mathcal M}^{ab}_I\psi)$.
Inserting these expressions in (\ref{YH})  and using the curved PCO \eqref{YFD} we are left with 
 \begin{eqnarray}
\label{YNnew}
S &&=\int_{\mathcal{SM}^{(4|4)}} {\mathcal L}^{(4|0)} \wedge \mathbb{Y}^{(0|4)}  \nonumber  \\ 
&&= \int_{\mathcal{SM}^{(4|4)}} \left( {\mathcal L}^{I}  (\bar\psi {\mathcal M}_{I,ab}\psi) V^a V^b 
\right) \wedge \mathbb{Y}^{(4|0)} \\
&& = \int_{_{\mathcal{SM}^{(4|4)}}} \!\!\left( \epsilon^{abcd}  
(F_{ab} \bar\iota \gamma_{cd} \iota T +  \chi_a \gamma_{bc} \iota T_a) \bar\Theta \Theta 
+i 
(F_{ab} \bar\iota \gamma_5\gamma_{cd} \iota T +  \chi_a \gamma_5\gamma_{bc} \iota T_a)
 \bar\Theta \gamma_5 \Theta 
\right) V^4 \delta^4(\psi) \nonumber 
\end{eqnarray}
The additional superfields 
$\bar\iota \gamma_{cd} \iota T$ and $ \iota_{\alpha} T_a$ are defined in terms of the 
2-form potential $T = T_{ab} V^a V^b + T_{\alpha a} V^a \psi^\alpha + T_{\alpha\beta} \psi^\alpha \psi^\beta$ 
\begin{eqnarray}
\label{YQnew}
\bar\iota \gamma_{cd} \iota T = \gamma^{\alpha\beta}_{cd} T_{\alpha\beta}\,, 
~~~~~~
\iota_{\alpha} T_a = T_{\alpha a}
\end{eqnarray}
Integrating on the cotangent space we are left with the integral on the bosonic  $x^a$ and on the fermionic $\theta^\alpha$ coordinates:  
\begin{eqnarray}
\label{YOnew}
S = \int \!\!\left(    E
\epsilon^{abcd} (F_{ab} \bar\iota \gamma_{cd} \iota T +  \chi_a \gamma_{bc} \iota T_a) \bar\Theta \Theta 
+i 
E\epsilon^{abcd} (F_{ab} \bar\iota \gamma_5\gamma_{cd} \iota T +  \chi_a \gamma_5\gamma_{bc} \iota T_a)
 \bar\Theta \gamma_5 \Theta 
\right) 
\end{eqnarray}
where $F_{ab}, \chi_a$ 
are the superfields appearing in the parametrization of $F$ (see \eqref{leaA}). 
The superfield $f_a$ is absent due to the Dirac matrix structure of the PCO that 
projects out the Dirac structure corresponding to $f_a$. Notice that at the end we have to compute the 
Berezin integral over the 
fermionic coordinates, and this selects inside the superfields $T_{\alpha\beta}, 
T_{\alpha a}, F_{ab}$ and $\chi_a$ all the needed pieces for reconstructing the supergravity 
in the new minimal formulation. Again, we notice that it is the auxiliary field term that reproduces 
the entire supergravity action. 

To compare the final action with the superspace action, one needs to identify 
all components of the superfields in \eqref{YQnew} and then perform the 
integration. We do not present this computation here; for this we refer to 
\cite{Grimm:1984pj,Girardi:1984eq,Muller:1985vga} 
and to the review \cite{Muller:1988ux}. This concludes the proof of the equivalence between 
the group-manifold approach and the superspace approach to new minimal $D=4$ supergravity. 

\section*{Acknowlegements}

We would like to thank our friends and collaborators:  L. Andrianopoli,  R. Catenacci, C.A. Cremonini, 
R. D'Auria, O. Hulik, S. Noja, R. Norris, L. Ravera, G. Tartaglino-Mazzucchelli and M. Trigiante, for useful discussions and comments. The work is partially funded by the University of Eastern Piedmont with FAR-2019 projects. P.A.G. thanks the Simons Center for Geometry and Physics (Stony Brook University, NY), 
where the present work has been completed. 

\vspace{1cm}
\appendix
\noindent
{\bf \Large Appendices}
\section{Forms on Supermanifolds}

We collect here some basic definitions and facts about integration on supermanifolds and on \emph{integral forms}. For exhaustive introductions to integral forms, we refer the reader to \cite{Witten:2012bg,Noja}, while for their use in physics, we refer to \cite{Castellani:2014goa, Castellani:2015paa, Cremonini:2020skt}.

Given a (smooth) supermanifold $\mathcal{M}^{(4|4)}$, the cotangent space $\mathcal{T}^*_P \mathcal{M}^{(4|4)}$ at a given point $P \in \mathcal{M}^{(4|4)}$ has both an even and an odd part, generated, in a given system of local coordinates $\left( x^a , \theta^\alpha \right), i=a,\ldots,4 , \alpha=1,\ldots,4$, by the $(1|0)$-forms $\left\lbrace d x^a , d \theta^\alpha \right\rbrace$, called \emph{superforms}, which are respectively odd and even. They have the following (super)commuting properties: 
\begin{equation}\label{IFA}
	d x^a \wedge d x^b = - d x^b \wedge d x^a \ , \ d  \theta^\alpha \wedge d  \theta^\beta = d  
	\theta^\beta \wedge d  \theta^\alpha \ , \ d x^a \wedge d  \theta^\alpha = {-}d \theta^\alpha \wedge d x^a \ .
\end{equation}
A generic $(p|0)$-form is an object of the (graded)symmetric power of $\mathcal{T}^*_P \mathcal{M}^{(4|4)}$ and it locally reads as
\begin{equation}\label{IFB}
	\omega^{(p|0)} = \omega_{[a_1 \ldots a_r](\alpha_1 \ldots \alpha_s)} \left( x , \theta \right) d x^{a_1} \wedge \ldots \wedge d x^{a_r} \wedge d  \theta^{\alpha_1} \wedge \ldots \wedge d  \theta^{\alpha_s} \ , \ p=r+s \ ,
\end{equation}
where the coefficients $\omega_{[a_1 \ldots a_r](\alpha_1 \ldots \alpha_s)}(x,\theta)$ are a set of superfields and the indices $a_1 \dots a_r$, $\alpha_1 \dots \alpha_s$ are antisymmetrized and symmetrized, respectively, due to \eqref{IFA}. The analog of the determinant bundle can be found in a different form complex, the complex of \emph{integral forms}. One can introduce the \emph{Berezinian bundle} $\mathpzc{B}er \left( \mathcal{M}^{(4|4)} \right)$, i.e., the space of objects which transform as the Berezinian (i.e., the \emph{superdeterminant}) under coordinate transformations. Integral forms are then constructed on open sets starting from this space and tensoring with (graded)symmetric powers of the parity-changed tangent space (see the recent review \cite{Noja} for a rigorous introduction to the subject). A practical and computationally powerful realization of the Berezinian and integral forms is given in terms of (formal) Dirac distributions on the cotangent space; a generic $(p|4)$-integral form can be locally described as
\begin{equation}\label{IFC}
	\omega^{(p|N)} = \omega_{[a_1 \ldots a_r]}^{(\alpha_1 \ldots \alpha_s)} \left( x , \theta \right) d x^{a_1} \wedge \ldots \wedge d x^{a_r} \wedge \iota_{\alpha_1} \ldots \iota_{\alpha_s} \delta \left( d  \theta^1 \right) \wedge \ldots \wedge \delta \left( d  \theta^4 \right) \ , \ p=r-s \ ,
\end{equation}
and the second number of the $(p|4)$-form keeps track of the number of Dirac deltas and is called \emph{Picture number}. The contraction $\iota_{\alpha}$ is defined as $\iota_{\alpha} \equiv \partial/\partial \psi^{\alpha}$. The formal Dirac deltas satisfy the following properties:
\begin{align}\label{IFD}\nonumber
	\int_{d  \theta} \delta \left( d  \theta \right) &= 1, \quad d  \theta \wedge \delta \left( d  \theta \right) = 0, \quad \delta \Big( d  \theta^\alpha \Big) \wedge \delta \left( d  \theta^\beta \right) = - \delta \left( d  \theta^\beta \right) \wedge \delta \Big( d  \theta^\alpha \Big), \\ 
	d x \wedge \delta \left( d  \theta \right) &={+} \delta \left( d  \theta \right) \wedge d  x,\quad
	\delta \left( \lambda d  \theta \right) = \frac{1}{\lambda} \delta \left( d  \theta \right), \quad d  \theta \wedge \iota^p \delta \left( d  \theta \right) = - p \iota^{p-1} \delta \left( d  \theta \right) \ \ .
\end{align}
The first property defines how $\delta \left( d  \theta \right)$'s have to be used to perform form integration along the commuting directions $d  \theta$'s; the second property reflects the usual property of the support of the Dirac distribution; the third and fourth properties imply $\delta \left( d  \theta \right)$'s are odd objects and together with the fifth property they indicate that actually, these are not distributions, but rather \emph{de Rham currents}, i.e., they define an \emph{oriented} integration; the last property allows the usual integration by parts of the Dirac delta.

A top form reads as
\begin{equation}\label{IFE}
	\omega_{top}^{(4|4)} \equiv \omega^{(4|4)} = \omega \left( x , \theta \right) \epsilon_{a_1 \ldots a_4} d x^{a_1} \wedge \ldots \wedge d x^{a_4} \wedge \epsilon_{\alpha_1 \ldots \alpha_4} \delta \left( d  \theta^{\alpha_1} \right) \wedge \ldots \wedge \delta \left( d  \theta^{\alpha_4} \right) \ ,
\end{equation}
where $\omega \left( x , \theta \right)$ is a superfield. Any integral form of any form degree $p$ can be obtained by acting with $4-p$ contractions on \eqref{IFE}.

One can also consider other classes of forms, called \emph{pseudoforms}, with a non-maximal and non-zero number of deltas. A general pseudoform with $q$ deltas is locally given by
\begin{equation}\label{IFH}
	\omega^{(p|q)} = \omega_{[a_1 \ldots a_r](\alpha_1 \ldots \alpha_s)[\beta_1 \ldots \beta_q]} \left( x , \theta \right) d x^{a_1} \wedge \ldots \wedge d x^{a_r} \wedge d  \theta^{\alpha_1} \wedge \ldots \wedge d  \theta^{\alpha_s} \wedge \delta^{(t_1)} \left( d  \theta^{\beta_1} \right) \wedge \ldots \wedge \delta^{(t_q)} \left( d  \theta^{\beta_q} \right) \ ,
\end{equation}
where we used the compact notation $\delta^{(t)} \left( d  \theta \right) \equiv \left( \iota \right)^t \delta \left( d  \theta \right)$. The form number is obtained as 
\begin{equation}\label{IFI}
	p = r + s - \sum_{i=1}^q t_i \ ,
\end{equation}

If $q=0$ we have superforms, if $q=4$ we have integral forms and if $0<q<4$ we have pseudoforms. These kinds of forms are to be used for example in \cite{Cremonini:2020skt} to construct objects which implement naturally the self-duality condition on supermanifolds. This is a consequence of the fact that the Hodge operator 
 \cite{Castellani:2014goa, Castellani:2015paa} on supermanifolds changes not only the form number but also the picture number:
\begin{equation}\label{IFJ}
	\star: \Omega^{(p|q)} \left( \mathcal{M}^{(4|4)} \right) \to \Omega^{(4-p|4-q)} \left( \mathcal{M}^{(4|4)} \right) \ .
\end{equation}
A notable example of an integral form is the \emph{Picture Changing Operator}: 
it is a $(0|4)$-form, in the cohomology of the operator $d $. It is used to lift 
a superform to an integral form by multiplication:
\begin{eqnarray}
	\nonumber \mathbb{Y}^{(0|4)} : \Omega^{(p|0)} \left( \mathcal{M}^{(4|4)} \right) &\to& \Omega^{(p|4)} \left( \mathcal{M}^{(4|4)} \right) \\
	\omega^{(p|0)} &\mapsto& \omega^{(p|4)} = \omega^{(p|0)} \wedge \mathbb{Y}^{(0|4)} \ .
\end{eqnarray}

\section{Curved Superspace} 

\def\fK{ \stackrel{\leftarrow}{K} } 

One interesting formula in \cite{Kuzenko:1998tsq,GGRS} used in the context of superspace supergravity gives 
the right action of an even vector field $K$ on a superfield. It is a complicated formula, albeit 
useful and important,  and can be translated into our language using the Hodge dual operator. We hope that our re-formulation might add some new insights to the original work in  \cite{GGRS,Kuzenko:1998tsq} 
from a different perspective. 

We consider the following definition 
\begin{eqnarray}
\label{KB}
\phi \cdot {\fK} = (-1)^M  \partial_M (\phi K^M)   
\end{eqnarray}
for an even superfield $\phi$ and an even vector $K$. This 
 also implies 
\begin{eqnarray}
\label{KA}
\phi \cdot e^{\fK} = (1\cdot e^{\fK}) e^K \phi
\end{eqnarray}
where $\phi$ is a super field. 

We would like to translate (\ref{KB}) using our Hodge dual operator. For that we observe that (\ref{KB}) corresponds to 
the action of a vector field on a density, which can be obtained by acting with the Hodge dual of a Lie derivative on a top form.  Then, using 
the Hodge dual operator $\star$ we set  
\begin{eqnarray}
\label{KC} 
\phi \cdot {\fK}  = {\mathcal L}_K^\dagger \phi \equiv \star   {\mathcal L}_K (\star \phi) 
\end{eqnarray}
where $\star \phi = \phi {\rm Vol}^{(4|4)}$. The equation can be easily verified using 
the properties of ${\rm Vol}^{(4|4)}$ and of the Lie derivative ${\mathcal L}_K$.
The derivation of (\ref{KA}) is now straightforward 
\begin{eqnarray}
\label{KD}
\phi \cdot e^{\fK} &=& \left( e^{{\mathcal L}_K} \right)^\dagger \phi = 
\star e^{{\mathcal L}_K} ( \star \phi) = 
\star e^{{\mathcal L}_K} ( \star 1  \, \phi) = 
  \star \left( e^{{\mathcal L}_K}  (\star 1) \right)  e^{{\mathcal L}_K}  \phi \nonumber \\
 &=& \left(  \star e^{{\mathcal L}_K}   {\rm Vol}^{(4|4)} \right)  e^{{\mathcal L}_K}  = (1\cdot e^{\fK}) e^K \phi
\end{eqnarray}
where $\star 1 = {\rm Vol}^{(4|4)}$ and 
\begin{eqnarray}
\label{KG} 
(1\cdot e^{\fK}) = (e^{{\mathcal L}_K})^\dagger 1 =  \star   e^{{\mathcal L}_K} \omega^{(4|4)} = \star
\left( {\rm Sdet}(J) \omega^{(4|4)}\right) = {\rm Sdet}(J)
\end{eqnarray}
Another interesting formula is 
\begin{eqnarray}
\label{KE}
(1\cdot e^{\fK})^{-1} = e^K (1\cdot e^{-\fK})
\end{eqnarray}
which can be proven as follows 
\begin{eqnarray}
\label{KF}
(1\cdot e^{-\fK}) \cdot e^{\fK} =  \left( e^{{\mathcal L}_K} \right)^\dagger  \left( e^{-{\mathcal L}_K} \right)^\dagger 1 = 1
\end{eqnarray}
therefore the superfield $(1\cdot e^{-\fK})$ is the inverse of the operator $\cdot e^{\fK}$, therefore it is 
the inverse of $ (1\cdot e^{\fK}) e^K $. Analogously, $(1\cdot e^{\fK})$ is the inverse of $e^K (1\cdot e^{-\fK})$.  

This is only an example of how the use of the Hodge dual operator simplifies several complicated computations in superspace. We believe that this new point of view might prove to be useful in the superspace formulation of supergravity.  

 \section{The Chiral Projectors using Picture Lowering Operators}

 Here we want to show the correspondence between the geometric formulation of $D=4, N=1$ old minimal supergravity, via the super-Hodge operator, with its superspace formulation using the Picture Lowering Operators. 
In particular, we have seen that the action can be written as
\begin{equation}\label{OSPE}
	\int_{\mathcal{SM}^{(4|4)}} \mathcal{L}^{(4|4)} = \int_{\mathcal{SM}^{(4|4)}} \star 1 = \int_{\mathcal{SM}^{(4|2)}_{chiral}} \mathcal{R}_{L} {\rm Vol}_L^{(4|2)} + \int_{\mathcal{SM}^{(4|2)}_{anti-chiral}}{\mathcal{R}}_{R} {\rm Vol}_R^{(4|2)} \ .
\end{equation}
We can obtain a $(4|2)$-form out of a $(4|4)$-form by decreasing the picture number by two. The geometrical operator that decreases the picture number is given by
\begin{equation}\label{OSPF}
	\mathbb{Z}^{(0|-1)}_X = \left\lbrace d , -i \Theta \left( \iota_X \right) \right\rbrace \ ,
\end{equation}
where $X$ is an odd vector field and the (odd) operator $\Theta$ is defined via its Fourier integral representation:
\begin{equation}\label{OSPG}
	\Theta \left( \iota_X \right) \equiv - i \lim_{\epsilon \to 0} \int_{- \infty} ^\infty \frac{dt}{t + i \epsilon} e^{i t \iota_X} \ .
\end{equation}
The key point about the operator $\mathbb{Z}$ is that it is a quasi-isomorphism, i.e., it maps cohomology classes into cohomology classes. This means that it allows to obtain a form with the same degrees of freedom but with a different picture number, out of a known form. This is strictly related to the same properties of the picture-changing operator $\mathbb{Y}$ described in the previous Sections. Indeed, $\mathbb{Y}$ represents a quasi-isomorphism and is related to $\mathbb{Z}$ via a quasi-inverse relation (i.e., they are inverse on cohomology classes), which can be schematically represented as
\begin{equation}\label{OSPH}
	\mathbb{Z} \mathbb{Y} \mathbb{Z} = \mathbb{Z} \ , \ \mathbb{Y} \mathbb{Z} \mathbb{Y} = \mathbb{Y} \ .
\end{equation}
In the case under consideration, we can project from the whole supermanifold to the chiral/anti-chiral submanifolds using the following operator:
\begin{equation}\label{OSPI}
	\mathbb{Z}^{(0|-2)} = \mathbb{Z}_C^{(0|-2)} + \bar{\mathbb{Z}}_{\bar C}^{(0|-2)} = \bar{\mathbb{Z}}_{\bar{\nabla}_{\dot{2}}} \bar{\mathbb{Z}}_{\bar{\nabla}_{\dot{1}}} + \mathbb{Z}_{\nabla_2} \mathbb{Z}_{\nabla_1} \ ,
\end{equation}
Thus, $\mathbb{Z}_c^{(0|-2)}$ is made out of the two vector fields $\bar{\nabla}_{\dot{\alpha}}$ and 
$\bar{\mathbb{Z}}_{ac}^{(0|-2)}$ is made out of the two vector fields $\nabla_{\alpha}$.\footnote{In the present section we use the Weyl/anti-Weyl notation to simplify the computations, see \cite{WB} for the conventions.} We will discuss the action of $\mathbb{Z}_c^{(0|-2)}$ only, the anti-chiral calculation being similar. Let us start by applying the first operator $\bar{\mathbb{Z}}_{\bar{\nabla}_{\dot{1}}}$ to the Lagrangian $\mathcal{L}^{(4|4)} = \star 1 = V^a V^b V^c V^d \epsilon_{abcd} \delta \left( \psi^\alpha \right) \epsilon_{\alpha \beta} \delta \left( \psi^\beta \right) \delta \left( \bar{\psi}^{\dot{\alpha}} \right) \epsilon_{\dot{\alpha} \dot{\beta}} \delta \left( \bar{\psi}^{\dot{\beta}} \right)$; first of all, we notice that since $\mathcal{L}^{(4|4)}$ is a top form, its exterior derivative is trivially zero, hence the action of the operator $\bar{\mathbb{Z}}_{\bar{\nabla}_{\dot{1}}}$ reads
\begin{eqnarray}
	\nonumber \bar{\mathbb{Z}}_{\bar{\nabla}_{\dot{1}}} \mathcal{L}^{(4|4)} &=& d \Theta \left( \bar{\iota}_{\bar{\nabla}_{\dot{\alpha}}} \right) V^a V^b V^c V^d \epsilon_{abcd} \delta \left( \psi^\alpha \right) \epsilon_{\alpha \beta} \delta \left( \psi^\beta \right) \delta \left( \bar{\psi}^{\dot{\alpha}} \right) \epsilon_{\dot{\alpha} \dot{\beta}} \delta \left( \bar{\psi}^{\dot{\beta}} \right) = \\
	\label{OSPJ} &=& d \left[ {\rm Vol}_L^{(4|2)} \frac{1}{\bar{\psi}^{\dot{1}}} \delta \left( \bar{\psi}^{\dot{2}} \right) \right] \ .
\end{eqnarray}
At this point, it is worth observing that the intermediate step of calculation involves \emph{inverse forms}, i.e., forms of the type $1/\psi$. For details of their geometrical interpretation and string-theoretic counterpart, we refer the reader to \cite{CGN}. Here, we only note that inverse forms emerge in intermediate passages, but disappear at the end of calculations. In \eqref{OSPJ} we collected the $V$'s and the undotted $\psi$'s in the chiral volume form ${\rm Vol}_L^{(4|2)}$ (see \eqref{CVOFA}); notice that this volume form  is $d$-closed (the same holds for the anti-chiral one): since the expression is covariant, we can substitute the action of the differential $d$ with the action of the covariant one, defined in \eqref{RasuperPoincare}-\eqref{rhosuperPoincare}. The closure is directly verified: the covariant derivative on $V$'s either involves the torsion $R^a$, which is zero or involves $\psi \bar{\psi}$, hence giving zero because of the presence of the two undotted delta's $\delta \left( \psi^\alpha \right) \epsilon_{\alpha \beta} \delta \left( \psi^\beta \right)$; the covariant derivative on $\psi$'s is the curvature $\rho$, which is proportional to at least one $V$, which is then annihilated by the presence of the other four $V$'s. Hence \eqref{OSPJ} becomes
\begin{equation}\label{OSPK}
	\bar{\mathbb{Z}}_{\bar{\nabla}_{\dot{1}}} \mathcal{L}^{(4|4)} = {\rm Vol}_L^{(4|2)} d \left[ \frac{1}{\bar{\psi}^{\dot{1}}} \delta \left( \bar{\psi}^{\dot{2}} \right) \right] \ .
\end{equation}
Now the key point is that the expression between square brackets is not covariant, hence the action of the exterior derivative involves the spin connection. In particular, we have
\begin{eqnarray}
	\nonumber \hspace{-0.5cm} {\rm Vol}_L^{(4|2)} d \left[ \frac{1}{\bar{\psi}^{\dot{1}}} \delta \left( \bar{\psi}^{\dot{2}} \right) \right] &=& {\rm Vol}_L^{(4|2)} \left[ -\frac{1}{\left(\bar{\psi}^{\dot{1}} \right)^2} \left( \bar{\rho}^{\dot{1}} + \frac{1}{4} \omega^{ab} \left( \sigma_{ab} \bar{\psi} \right)^{\dot{1}} \right) \delta \left( \bar{\psi}^{\dot{2}} \right) + \right. \\
	 \label{OSPL} &+& \left. \frac{1}{\bar{\psi}^{\dot{1}}} \left( \bar{\rho}^{\dot{2}} + \frac{1}{4} \omega^{ab} \left( \sigma_{ab} \bar{\psi} \right)^{\dot{2}} \right) \delta^{(1)} \left( \bar{\psi}^{\dot{2}} \right) \right] = \\
	\nonumber &=& {\rm Vol}_L^{(4|2)} \left[ \frac{-1}{4 \left(\bar{\psi}^{\dot{1}} \right)^2} \omega^{ab} \left( \sigma_{ab} \bar{\psi} \right)^{\dot{1}} \delta \left( \bar{\psi}^{\dot{2}} \right) + \frac{1}{4 \bar{\psi}^{\dot{1}}} \omega^{ab} \left( \sigma_{ab} \bar{\psi} \right)^{\dot{2}} \right] \delta^{(1)} \left( \bar{\psi}^{\dot{2}} \right) \ ,
\end{eqnarray}
where the terms involving $\rho$ have been dropped, since they contain at least one vielbein $V$.  We use the 
notation 
\begin{equation}\label{OSPM}
	- \left( \sigma_{ab} \bar{\psi} \right)^{\dot{\alpha}} = \left[ P_{-} \left( \gamma_{ab} \psi \right) \right]^{\dot{\alpha}} = \left[ \left( \gamma_{ab} P_{-} \psi \right) \right]^{\dot{\alpha}} \ , \ P_{-} = \frac{1-\gamma_5}{2} \ , $$ $$ \sigma_{a ~ \dot{\beta}}^{~ \alpha} = \left( 1 , \sigma_i \right)^\alpha_{~ \dot{\beta}} \ , \ \sigma_{a ~ \beta}^{~ \dot{\alpha}} = \left( 1 , -\sigma_i \right)^{\dot{\alpha}}_{~ \beta} \ , \ \gamma_a = \left( \begin{matrix}
		0 & \sigma_{a ~ \dot{\beta}}^{~ \alpha} \\
		- \sigma_{a ~ \beta}^{~ \dot{\alpha}} & 0
	\end{matrix} \right) \ ,
\end{equation}
$\sigma_i$ being the Pauli matrices. We could move to the more convenient chiral notation for the action of Lorentz generators:
\begin{equation}\label{OSPN}
	\omega^{ab} \left( \sigma_{ab} \psi \right)^{\alpha} = \omega^\alpha_{~ \beta} \psi^\beta \ , \ \omega^{ab} \left( \sigma_{ab} \bar{\psi} \right)^{\dot{\alpha}} = \omega^{\dot{\alpha}}_{~ \dot{\beta}} \psi^{\dot{\beta}} \ , \ \omega^{\alpha \beta} = \omega^{\beta \alpha} \ , \ \omega^{\dot{\alpha} \dot{\beta}} = \omega^{\dot{\beta} \alpha} \ .
\end{equation}
Eq. \eqref{OSPL} then becomes
\begin{eqnarray}
	\nonumber \bar{\mathbb{Z}}_{\bar{\nabla}_{\dot{1}}} \mathcal{L}^{(4|4)} &=& {\rm Vol}_L^{(4|2)} \left[ \frac{-1}{4 \left(\bar{\psi}^{\dot{1}} \right)^2} \left( \omega^{\dot{1}}_{~ \dot{1}} \bar{\psi}^{\dot{1}} + \omega^{\dot{1}}_{~ \dot{2}} \bar{\psi}^{\dot{2}} \right) \delta \left( \bar{\psi}^{\dot{2}} \right) + \frac{1}{4 \bar{\psi}^{\dot{1}}} \left( \omega^{\dot{2}}_{~ \dot{1}} \bar{\psi}^{\dot{1}} + \omega^{\dot{2}}_{~ \dot{2}} \bar{\psi}^{\dot{2}} \right) \right] \delta^{(1)} \left( \bar{\psi}^{\dot{2}} \right) = \\
	\nonumber &=& - {\rm Vol}_L^{(4|2)} \left[ \frac{\omega^{\dot{1}}_{~ \dot{1}}}{4 \bar{\psi}^{\dot{1}}} + \frac{\omega^{\dot{2}}_{~ \dot{2}}}{4 \bar{\psi}^{\dot{1}}} \right] \delta \left( \bar{\psi}^{\dot{2}} \right) + \frac{1}{4} \omega_c^{(4|2)}  \omega^{\dot{2}}_{~ \dot{1}} \delta^{(1)} \left( \bar{\psi}^{\dot{2}} \right) = \\
	\label{OSPO} &=& \frac{1}{4} {\rm Vol}_L^{(4|2)}  \omega^{\dot{2}}_{~ \dot{1}} \delta^{(1)} \left( \bar{\psi}^{\dot{2}} \right) \ .
\end{eqnarray}
Notice that the terms containing inverse forms have canceled, as expected, because of the relation $\omega^{\dot{1}}_{\dot{1}} = - \omega^{\dot{2}}_{\dot{2}} $. We can now act with the second PCO $\bar{\mathbb{Z}}_{\nabla_{\dot{2}}}$; since $\bar{\mathbb{Z}}$ is a cohomology operator, its action can be simplified as
\begin{equation}\label{OSPP}
	\bar{\mathbb{Z}}_{\bar{\nabla}_{\dot{2}}} \bar{\mathbb{Z}}_{\bar{\nabla}_{\dot{1}}} \mathcal{L}^{(4|4)} = d \Theta \left( \iota_{\bar{\nabla}_{\dot{2}}} \right) \left[ \frac{1}{4} {\rm Vol}_L^{(4|2)} \omega^{\dot{2}}_{~ \dot{1}} \delta^{(1)} \left( \bar{\psi}^{\dot{2}} \right) \right] = - \frac{1}{4} {\rm Vol}_L^{(4|2)} d \left[ \omega^{\dot{2}}_{~ \dot{1}} \frac{1}{\left( \bar{\psi}^{\dot{2}} \right)^2} \right] \ .
\end{equation}
Again, inverse forms arise, but they cancel after the action of the exterior derivative. In particular, we have
\begin{eqnarray}
	\nonumber - \frac{1}{4} {\rm Vol}_L^{(4|2)} d \left[ \omega^{\dot{2}}_{~ \dot{1}} \frac{1}{\left( \bar{\psi}^{\dot{2}} \right)^2} \right] &=& - \frac{1}{4} {\rm Vol}_L^{(4|2)} \left[ d \omega^{\dot{2}}_{~ \dot{1}} \frac{1}{\left( \bar{\psi}^{\dot{2}} \right)^2} + 2 \omega^{\dot{2}}_{~ \dot{1}} \frac{1}{\left( \bar{\psi}^{\dot{2}} \right)^3} \left( \omega^{\dot{2}}_{~ \dot{1}} \bar{\psi}^{\dot{1}} + \omega^{\dot{2}}_{~ \dot{2}} \bar{\psi}^{\dot{2}} \right) \right] = \\
	\label{OSPQ} &=& - \frac{1}{4} {\rm Vol}_L^{(4|2)} \frac{1}{\left( \bar{\psi}^{\dot{2}} \right)^2} \left[ d \omega^{\dot{2}}_{~ \dot{1}} + 2 \omega^{\dot{2}}_{~ \dot{1}} \omega^{\dot{2}}_{~ \dot{2}} \right] \ .
\end{eqnarray}
The expression in square brackets represents exactly a component of the curvature \eqref{RabsuperPoincare} in the chiral/antichiral notation. The terms containing $V$'s and $\psi$'s are automatically annihilated by the presence of $\omega_c^{(4|2)}$, while only after some algebra it is possible to verify that only the $\left( \bar{\psi}^{\dot{2}} \right)^2$ term survives. In particular, we arrive at the final expression (up to an overall factor)
\begin{equation}\label{OSPR} 
	\bar{\mathbb{Z}}_{\bar{\nabla}_{\dot{2}}} \bar{\mathbb{Z}}_{\bar{\nabla}_{\dot{1}}} \mathcal{L}^{(4|4)} = {\rm Vol}_L^{(4|2)} \left( S + i P \right) = {\rm Vol}_L^{(4|2)} \mathcal{R}_{L} \ .
\end{equation}
Analogous computations can be performed by acting on $\mathcal{L}^{(4|4)}$ with the Picture Lowering Operators along the $\nabla_{\alpha}$ directions, so that finally one obtains
\begin{equation}\label{OSPS} 
	\left( \bar{\mathbb{Z}}_{\bar{\nabla}_{\dot{2}}} \bar{\mathbb{Z}}_{\bar{\nabla}_{\dot{1}}} + \mathbb{Z}_{\nabla_2} \mathbb{Z}_{\nabla_1} \right) \mathcal{L}^{(4|4)} = {\rm Vol}_L^{(4|2)} \mathcal{R}_{L} + {\rm Vol}_R^{(4|2)} {\mathcal{R}}_{R} \ ,
\end{equation}
reproducing the usual action \eqref{OSPE}.

\section{Some Formulas for Supergravity}
In this Appendix, we give some explicit formulas used in the text and their Hodge dual.
Using supergravity parametrizations we have 
\begin{eqnarray}
\label{TorDualA}
T^a &=& 0\nonumber \\
\rho^\alpha &=& \rho^\alpha_{bc} V^b \wedge V^c + 
\rho^\alpha_{\beta c} \psi^\beta \wedge V^c \,, 
\end{eqnarray}
with $\rho^\alpha_{\beta c}  = 
i \Big(\gamma_{a} S  - i \gamma_a \gamma_5 P + \gamma_5 (2 \delta^b_a -   \gamma^b_a) A_b \Big)^\alpha_{~\beta}$ (where I set $c_4 =1$) 
Then, 
\begin{eqnarray}
\label{TorDualA}
\star T^a &=& 0\nonumber \\
\star \rho^\alpha &=& \rho^\alpha_{bc}  V_2^{de} \delta^4(\psi) + 
\rho^\alpha_{c \beta}  C^{\beta\gamma}V^{c}_3 \iota_\gamma \delta^4(\psi)\,, 
\end{eqnarray}
Now acting with the differential $d$, we finally get 
\begin{eqnarray}
\label{TorDualA}
\nabla \star T^a &=& 0 \\
\nabla \star \rho^\alpha &=& \nabla_b \rho^\alpha_{cd} \epsilon^{bcda'} \eta_{a'a}  
V_3^{a} \delta^4(\psi) + \left( \eta^{bc} \nabla_b \rho^\alpha_{c \beta}
+ \rho^\alpha_{ab} \rho^\gamma_{cd} \epsilon^{abcd} + \rho^\alpha_{c \beta} C^{\beta \rho} \rho^\gamma_{d \rho} \eta^{cb} 
\right) V_4 \iota_\gamma \delta^4(\psi) \nonumber 
\end{eqnarray}
and 
\begin{eqnarray}
\label{TorDualB}
\star \nabla \star \rho^\alpha = 
\nabla_b \rho^\alpha_{cd} \epsilon^{bcda'} \eta_{a'a} V^a + 
\left( \eta^{bc} \nabla_b \rho^\alpha_{c \beta}
+ \rho^\alpha_{ab} \rho^\gamma_{cd} \epsilon^{abcd} + \rho^\alpha_{c \beta} C^{\beta \rho} \rho^\gamma_{d \rho} \eta^{cb} 
\right) C_{\gamma \sigma} \psi^\sigma
\end{eqnarray}

It is very instructive to compute $d\star d\phi$ explicitly 
 using a generic supergravity 
 parametrization for the super vielbeins 
 \footnote{Following the conventions of \cite{book} we use 
the notation $T^\alpha = \rho^\alpha$. }
\begin{eqnarray}
\label{HDOD}
\nabla V^a = T^a &=& T^a_{bc} V^b \wedge V^c + T^a_{\beta c} \psi^\beta \wedge V^c + 
T^a_{\beta \gamma} \psi^\beta \wedge \psi^\gamma\,, \nonumber \\
\nabla \psi^\alpha = \rho^\alpha &=& \rho^\alpha_{bc} V^b \wedge V^c + 
\rho^\alpha_{\beta c} \psi^\beta \wedge V^c + 
\rho^\alpha_{\beta \gamma} \psi^\beta \wedge \psi^\gamma\,, 
\end{eqnarray}
where the superfields $ T^a_{bc}, \dots, \rho^\alpha_{\beta \gamma}$ 
are defined in terms of the superspace constraints (also known as {\it rheonomic constraints})  and using the Bianchi identities.
Then, we get
 \begin{eqnarray}
\label{HDOCAB}
&&d \star d \phi =\nonumber\\
&& \hspace{-.5cm} = d \left( V^{a}_3 \delta^4(\psi) \nabla_a \phi \right) + d
\left(C^{\alpha\beta} V^4 \iota_\alpha \delta^4(\psi) \nabla_\alpha \phi\ \right)  \nonumber \\
&&\hspace{-.5cm}= \frac12 \epsilon^{abcd} \nabla V_b V_c V_d \delta^4(\psi) \nabla_a \phi  
-  V^{a}_3 \nabla \psi^\alpha \iota_\alpha  \delta^4(\psi) \nabla_a \phi - 
V^{a}_3  \delta^4(\psi)  (V^d \nabla_d \nabla_a \phi + \psi^\delta \nabla_\delta \nabla_a \phi) \nonumber \\
&&\hspace{-.5cm}+  \frac{C^{\alpha\beta} }{3!} \epsilon^{abcd} \nabla V_a V_b V_c V_d \iota_\alpha\delta^4(\psi) \nabla_\beta\phi
+ C^{\alpha\beta} V^4 \nabla \psi^\gamma \iota_\gamma\iota_\alpha \delta^4(\psi) \nabla_\beta \phi 
+ C^{\alpha\beta} V^4 \iota_\alpha \delta^4(\psi) (V^d \nabla_d \nabla_\beta\phi + \psi^\delta \nabla_\delta \nabla_\beta\phi)
\nonumber 
\end{eqnarray}
 \begin{eqnarray}
\label{HDOCAC}
&&\hspace{-.5cm}= 
 \frac12 \epsilon^{abcd}  T_{b rs} V^r V^s V_c V_d \delta^4(\psi) \nabla_a \phi 
 -  V^{a}_3 (\rho^\alpha_{\beta c} \psi^\beta V^c) \iota_\alpha  \delta^4(\psi) \nabla_a \phi 
 - V^{a}_3  \delta^4(\psi)  V^d \nabla_d \nabla_a \phi \nonumber \\
 &&\hspace{-.5cm}+ 
 \frac{C^{\alpha\beta} }{3!} \epsilon^{abcd} T_{a \beta r} \psi^\beta V^r  V_b V_c V_d \iota_\alpha\delta^4(\psi) \nabla_\beta\phi 
 + C^{\alpha\beta} V^4 \rho^\alpha_{\beta \gamma} \psi^\beta \psi^\gamma 
 \iota_\gamma\iota_\alpha \delta^4(\psi) \nabla_\beta \phi 
+ C^{\alpha\beta} V^4 \iota_\alpha \delta^4(\psi) \psi^\delta \nabla_\delta \nabla_\beta\phi \nonumber \\
&&\hspace{-.5cm} =
V^4 \delta^4(\psi)  \Big[(T^{ab}_{a}  - \rho^{\alpha b}_\alpha) \nabla_b \phi + 
(T^{a\beta}_{a}  + \rho^{\alpha \beta}_\alpha) \nabla_\beta \phi -
(\eta^{ab} \nabla_a \nabla_b \phi + C^{\alpha\beta} \nabla_\alpha \nabla_\beta \phi) \Big]
\end{eqnarray}
where the coefficients in front of $\nabla_a\phi$ and $\nabla_\alpha\phi$ 
are obtained by suitable contraction of the supertorsion $T, \rho$.

Applying the Hodge dual directly on  the torsion, we get 
\begin{eqnarray}
\label{HDOE}
\star T^a &=& T^a_{bc} V_2^{bc} \delta^4(\psi)
 + T^a_{\beta c}   V^{c}_3 \iota_\alpha  \delta^4(\psi)
+ T^a_{\beta \gamma} V^4 \iota^\beta \iota^\gamma \delta^4(\psi)\,, \nonumber \\
\star\rho^\alpha &=& \rho^\alpha_{bc} V_2^{de} \delta^4(\psi)+ 
\rho^\alpha_{\beta c} V^{c}_3 \iota_\alpha  \delta^4(\psi)
 + 
\rho^\alpha_{\beta \gamma} V^4 \iota^\beta \iota^\gamma \delta^4(\psi)\,, 
\end{eqnarray}

\end{document}